\documentclass[twocolumn, 10pt, journal,letterpaper]{IEEEtran}
\IEEEoverridecommandlockouts
\usepackage{algorithm}
\usepackage{graphicx}
\usepackage[noend]{algpseudocode}
\usepackage{url} 
\usepackage{etex} 
\usepackage{tikz}
\usetikzlibrary{shapes,arrows}
\usepackage{pstricks}
\usepackage{pst-node,pst-blur}
\usetikzlibrary{arrows}
\usepackage{amsfonts}
\usepackage{tabularx}
\usepackage{enumerate}
\usepackage{subfigure}
\usepackage{amssymb}
\usepackage{steinmetz}   
\usepackage{booktabs}
\usepackage{multirow}
\usepackage{epsfig}
\usepackage{epstopdf}
\usepackage{array}

\usepackage[noadjust]{cite}
\newtheorem{Theorem}{Theorem}
\newtheorem{Remark}{Remark}
\newtheorem{Definition}{Definition}

\newtheorem{Lemma}{Lemma}

\newtheorem{Corollary}{Corollary}

\usepackage[cmex10]{amsmath}
\usepackage[cmex10]{mathtools}
\DeclarePairedDelimiter\ceil{\lceil}{\rceil}
\DeclarePairedDelimiter\floor{\lfloor}{\rfloor}
\algdef{SE}[DOWHILE]{Do}{doWhile}{\algorithmicdo}[1]{\algorithmicwhile\ #1}%
\makeatletter
\def\set@curr@file#1{%
  \begingroup
    \escapechar\m@ne
    \xdef\@curr@file{\expandafter\string\csname #1\endcsname}%
  \endgroup
}
\def\quote@name#1{"\quote@@name#1\@gobble""}
\def\quote@@name#1"{#1\quote@@name}
\def\unquote@name#1{\quote@@name#1\@gobble"}
\makeatother
\usepackage{graphics}

\allowdisplaybreaks  

\makeatletter 
\def\@eqnnum{{\normalsize \normalcolor (\theequation)}} 
\makeatother  
  
\begin{document}
\title{Utility-Aware Optimal Resource Allocation Protocol for  UAV-Assisted Small Cells with Heterogeneous Coverage Demands}

\author{Poonam Lohan,~\IEEEmembership{Student Member,~IEEE} and Deepak Mishra,~\IEEEmembership{Member,~IEEE}
\thanks{P. Lohan is with the Department of Electrical Engineering, Indian Institute of Technology Delhi, New Delhi 110016, India (e-mail: lohanp88@gmail.com).}
\thanks{D. Mishra is with the School of Electrical Engineering
and Telecommunications at the University of New South Wales (UNSW) Sydney, NSW 2052, Australia (email: d.mishra@unsw.edu.au).}
\thanks{ A preliminary version \cite{MISHRA} of this work has been published in the proceedings of IEEE ICC, Shanghai, China, May 2019.}}

\maketitle
\vspace{-5mm}
\begin{abstract}
In this paper, we consider a UAV-assisted small-cell having heterogeneous users with different data rate and coverage demands. Specifically, we propose a novel utility-aware resource-allocation protocol to maximize the utility of UAV by allowing it to simultaneously serve the highest possible number of heterogeneous users with available energy resources. In this regard, first we derive a closed-form expression for rate-coverage probability of a user considering Rician fading to incorporate the strong line of sight (LoS) component in UAV communication. Next since this UAV utility maximization problem is non-convex and combinatorial, to obtain the global optimal resource allocation policy we propose an iterative feasibility checking method for fixed integers ranging from lower to upper bound on the number of users that can be served by UAV. To further reduce the complexity, we formulate an equivalent problem aimed at minimizing per user energy consumption, where tight analytical relaxation on rate-coverage probability constraint is used along with semi-closed expressions for joint-optimal power and time allocation. Lastly, via detailed numerical investigation, we validate our analytical claims, present insights on the impact of key system parameters, and demonstrate that  $60 \%$ more users can be served using the proposed scheme as compared to relevant benchmarks. 
\end{abstract} 

\IEEEpeerreviewmaketitle
\bstctlcite{IEEEexample:BSTcontrol}

\section{Introduction}
Unmanned aerial vehicles (UAVs) have attracted great interest from the industry and academia to be utilized in serving a multitude of applications that include surveillance, monitoring, rescue operations, telecommunication, military, Internet of Things (IoT) communication \cite{Hayat}-\cite{BAEK}, and  public safety operations \cite{DIS}.  UAVs provide effective solution to serve temporary high traffic demands during high crowd events such as festivals, concerts, and stadium games. While UAVs  are very beneficial to provide fast and reliable communication services in the above discussed different scenarios, it is also important to use the available energy resources wisely such that highest possible number of users can be served by UAV, where users can have heterogeneity in their service demands in terms of different data rate and coverage requirements. To maximize the  UAV utility in terms of serving the highest  number of users in energy-efficient manner, proper  resource allocation to  heterogeneous users in accordance to their service requirement and channel condition  is very challenging task.
\subsection{Related Work} 
To get full advantage of inherent properties of UAVs which are not possible in conventional wireless networks such as flexible deployment, adaptive altitude, and line-of-sight (LoS) communication link to ground users, it is  important to address some technical challenges that includes coverage maximization, energy-efficiency, propagation characteristics, resource management, and performance analysis \cite{Asadpour}-\cite{YZeng}. 

 Many research works have focused on optimizing coverage performance of  UAV-assisted networks by finding the optimal placement and efficient  deployment strategies for UAVs  \cite{Hourani}-[18]. For example, an approach was proposed in \cite{Hourani} to find the optimal altitude  of a single static UAV to maximize its coverage area. In \cite{Moza_Glbcom}, the  authors considered two UAVs case with full interference between UAVs, and optimized the altitude and distance between UAVs to maximize the coverage area. In \cite{Yaliniz}-\cite{Alzenad1}, the authors proposed  an algorithm for optimal placement of UAV to maximize the number of users covered by UAV-assisted small cells.   By investigating a joint UAV trajectory and communication design, in \cite{QURUI} the authors characterize the capacity region of a new two-user broadcast channel with a UAV-mounted aerial base station, and in \cite{QURUI2} the authors maximize the minimum throughput over all ground users while considering a multiple UAV and multiple user system with TDMA for users and interference channel for UAVs.  An approach was discussed for optimal location of UAV to improve the throughput coverage gain for public safety communications in \cite{Merwa}. In \cite{Moza}-\cite{JLyu}, the authors proposed analytical and numerical method, respectively, to minimize the number of UAVs needed to provide coverage to a target area.  In \cite{Moza_IoT}, the authors studied efficient deployment of UAV, device UAV association, and uplink power control to minimize the energy while serving IoT devices. The work in \cite{Alzenad2} investigated the optimal  placement of UAV for maximizing the coverage of users having different QoS requirements in terms of signal to noise ratios.  The work in \cite{Moza_D2D} investigated the optimal placement of UAV to optimize the rate coverage  probability performance in device-to device-communication system. However, these works focused on maximizing coverage area while ignoring small-scale fading, did not discuss  how to optimally allocate the resources to the users under SNR coverage, such that the  maximum possible number of user  could achieve their required data rate. Some recent works also  considered small-scale fading  with  path loss in UAV communications for coverage analysis \cite{Azari,CZhang, Dhillon}.  In \cite{Azari}, the authors analyzed the performance of UAV-assisted communication in terms of outage probability while considering Rician fading and investigated the sum rate and power gain trade-off. Stochastic geometry based coverage probability analysis for a finite UAV wireless network over Rayleigh fading and Nakagami-m  channels was investigated in \cite{CZhang} and \cite{Dhillon} respectively.  However, considering small-scale fading these works did not provide closed-form expression for  coverage probability and ignored the underlying  rate-constrained resource allocation among users.

Another line of research has focused on  resource allocation and users scheduling in UAV-assisted  networks \cite{JLI,RFAN,Wang_joint,Soorki_M2M,QWu}.  The work in \cite{JLI} presented a framework  to minimize the mean packet transmission delay in multi-layer UAV network. In \cite{RFAN}, the authors studied joint optimization of UAV placement and resource allocation in terms of power and bandwidth, to maximize the  throughput for a UAV-relaying system. The work in \cite{Wang_joint} investigated the joint transmit power and trajectory optimization to maximize the minimum average throughput within a given time. In \cite{Soorki_M2M}, the authors proposed an optimal resource allocation and scheduling to minimize the transmission power consumption in UAV-assisted machine to machine communication scenario.  The delay-constrained communication  framework to maximize the  average throughput  by jointly optimizing the resource allocation and UAV trajectory was studied in \cite{QWu}.  However, in all these works~\cite{JLI,RFAN,Wang_joint,Soorki_M2M,QWu}, the  heterogeneity in service demands of users  was ignored.

\subsection{Novelty and Scope}
\emph{To the best of our knowledge, this is  first work that considers optimal resource allocation to maximize the number of users under service in UAV-assisted communication  while considering the heterogeneous users with different data rate and coverage requirement}. We present a novel resource allocation protocol and a novel analysis providing closed-form expression for rate-coverage probability over Rician channels. We propose a energy-efficient solution methodology having parallel computing for  non-convex combinatorial UAV utility maximization problem  providing closed-form joint-optimal solution for power and time allocation with very low complexity.

Although, existing works related to UAV communication focused on the efficient deployment, trajectory control, routing optimization, coverage area maximization, and UAV altitude optimization; optimal resource allocation can not be ignored where to serve multi-users, multiple access scheme is mandatory. Optimal resource allocation is fundamental requirement to tackle with the heterogeneity in service demands of users and  to provide services to as many number of users as possible with limited available energy resources.   In this work, we  consider single UAV  scenario, while the proposed solution methodology can be utilized in the multi-UAV scenario also where the problem can be
decomposed on per UAV utility maximization basis. Although, in multi-cell scenario, the utility maximization problem and the corresponding solution would be different and the detailed investigation is required with interference consideration which is out of scope of this work. Though we have considered the static UAV at a fixed altitude,  the applicability of the proposed solution methodology in mobile UAV scenario is discussed in Appendix C.

\subsection{Key Contribution and Paper Organization}
 The key contribution of this work is five-fold: (1) A novel resource allocation  protocol is considered  to maximize the UAV utility  by optimally allocating power and time resources to each user  (Fig$.$~\ref{fig:tran_proto}) (\emph{Section II}). (2) Next, the distribution of distance of randomly deployed users from UAV is derived and a closed-form expression for rate-coverage probability over Rician fading channels is obtained by using the tight exponential approximation of Marcum Q-function. Next, to gain more analytical insights, two special cases of high SNR regime and dominant LoS scenario are also considered  providing simpler closed-form expression for rate-coverage probability (\emph{Section III}). (3) An optimization problem is defined to maximize the UAV utility in terms of serving the highest possible number of heterogeneous users under energy resources and rate-coverage constraints. To consider the combinatorial aspect of problem, first, the generalized-convexity of the optimization problem is proved for a fixed integer value of number of users and tight analytical lower and upper bounds are provided on the number of servable users by UAV with available energy resources. Then, an iterative feasibility checking method is proposed which provide global optimal solution by checking the feasibility of Karush-Kuhn-Tucker (KKT) conditions of the problem for fixed integer value of number of users ranging from defined lower to upper bounds on it (\emph{Section IV}). (4) An equivalent  problem aiming at maximizing UAV utility by minimizing per user energy consumption is defined. Using a tight analytical relaxation on rate-coverage probability constraints, it is reduced to a single variable optimization problem. Furthermore, a joint optimization algorithm having very low solution complexity is proposed which provide closed-form solution for joint-optimal power and time allocation. Individual power and time optimization is also done for UAV utility maximization  (\emph{Section V}). (5) Numerical results validate the analysis, discuss design insights on optimal energy resource utilization, and compare the performance gain of the proposed scheme against the  benchmark fixed allocation scheme (\emph{Section VI}). Lastly, the paper is concluded in  \emph{Section VII}.

\section{System Model}
Here, we first present the system model including the  network topology and channel fading. Then, we outline the proposed resource allocation protocol for  all the heterogeneous users that are served by a  UAV deployed as aerial base station. 
\subsection{Network Topology and Channel Model}
\begin{figure}[pt] \centering
	\includegraphics[width=5.5 cm]{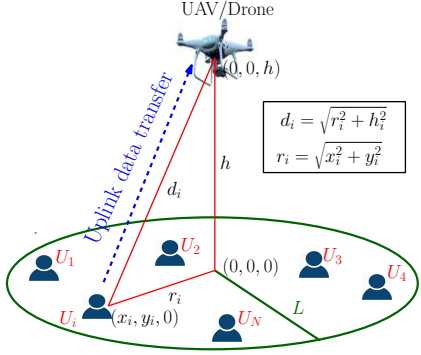} 
	\caption{System model showing UAV coverage area with $N$ users in 2D plane.}\label{fig:sys_mod}
\end{figure}

We consider a UAV-assisted uplink communication system, where $N$  users are uniformly distributed on the ground over the two dimensional circular field of radius $L$ meters (m). In particular as shown in  Fig$.$~\ref{fig:sys_mod}, these $N$ users  communicate with the single UAV. Without loss of generality, the  UAV is assumed to be static that hover at an altitude of $h$ meters above the center of circular field containing all the users. Hereinafter, each $i$-th user is denoted by $U_i,\; \forall i\in I{\triangleq}\{1,2,\ldots,N\}$. Let $(0,0,0)$, $(x_i,y_i,0)$, and  $(0,0,h)$ be the three-dimensional (3D) coordinates of center of circular field,  location of user $U_i ,\; \forall i\in I$, and UAV,  respectively. 
 Consider a tagged user $U_i$ as shown in Fig$.$~\ref{fig:sys_mod}, which is located at  distance $r_i\triangleq\sqrt{x_i^2+y_i^2}$, from the center of circular field. We have considered Euclidean distances in this work. Since, UAV is assumed to be located at an altitude of $h$ meters above the center of circular field, distance of user $U_i$ from UAV can be expressed as, $ d_i\triangleq \sqrt{r_i^2+h^2} $. The UAV and all users are assumed to be equipped with single antenna.
 
The wireless channels between  users and UAV are assumed to undergo statistically independent frequency non-selective quasi-static Rician block fading. Consequently, the random channel power gains, $ g_i,\; \forall i\in I $, are noncentral-$\chi^2$ distributed with mean $\bar{g_i}=\frac{\mu_i}{d_i^{\alpha}}$ and rice factor $K_i$ \cite{simon}. Here,  $\mu_i$ is the average channel power gain parameter that depends on antenna characteristics and average channel attenuation, $d_i$ is the user $U_i$ to UAV distance, and $\alpha$ is path loss exponent. The signal-to-noise ratio (SNR) $\gamma_i$ at UAV for user $U_i$ is written as: 
\begin{align}\label{eq:eq1}
\gamma_{i}\triangleq {P_i g_i }/{\sigma^2}, \; \forall i\in I,
\end{align}
where $P_i$ is the transmission power of user $U_i$ and $\sigma^2$ denotes the power of Additive White Gaussian Noise (AWGN). For reduced signaling overhead at all the users, instead of instantaneous channel state information (CSI), we assume the availability of statistics of CSI for all links at UAV. They are collected via pilot signals received from users. The rate coverage analysis and all optimization related computations are performed at UAV  using the statistical CSI.

\subsection{Proposed  Resource Allocation Protocol}
\begin{figure}[pt] \centering
\includegraphics[width=5.5 cm]{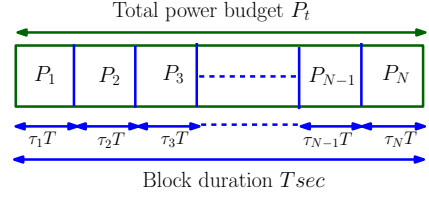}
\caption{Proposed resource allocation protocol.}\label{fig:tran_proto}
\end{figure}

We consider uplink scenario in which users transmit information to the UAV carried aerial base station. UAV adopts a time division multiple access technique (TDMA) to provide service to all the users.  It is assumed that all the users have to share a common power source with total budget $P_t$ Watt. The practical application of common power budget consideration include:
(1)  Multiple transmitters (TXs) powered by a single solar-panel-powered battery.
(2) Multiple sensor nodes are connected to the common power supply.
(3) Distributed antenna systems (DASs), where multiple antennas are geographically placed
at various locations and these antennas are connected to a common central source via wired
connections.  Let $P_i$ denotes the power allocated from power budget to user $U_i, \forall i\in I$. Since, we are considering TDMA, each user  transmits information to UAV  over the channel in a particular period of time that is allocated to the corresponding user. Let $\tau_i\in\{0,1\}$,  represent the  normalized fraction of time allocated from  block duration $T$ to user $U_i,\; \forall i\in I$ as shown in Fig$.$~\ref{fig:tran_proto}.  Using the Shannon's capacity formula and SNR representation in \eqref{eq:eq1}, the spectral efficiency in bits per sec per hertz (bps/Hz) for user $U_i$ to UAV communication link is given by:
 \begin{align}\label{eq:eq2}
\eta_i\triangleq \tau_i \log_2(1+\gamma_i) = \tau_i \log_2\left(1+{P_i g_i}/{\sigma^2}\right), \; \forall i\in I.
\end{align}
Now, we define the notion of a user being under rate-coverage by the UAV. This definition also will be used  for rate-coverage probability analysis in next section. 
\begin{Definition}A tagged user $U_i$ would be considered under UAV rate-coverage, if  spectral efficiency for that user is greater than its desired rate threshold $\eta_{th_i}$, i.e., $\eta_i\geq \eta_{th_i}, \; \forall i\in I$. 
\end{Definition}
Here our main aim is to maximize the UAV utility  by allocating power and time resources optimally to all the heterogeneous users. In other words, UAV should serve maximum  number of users with given energy resources while satisfying their  different  data rate and coverage demands. Note that, we are not  selecting users with lesser data rate and coverage demand to maximize  UAV utility, while  users are prioritized on the first come, first serve basis, i.e., user with lower index would be served first than the user with higher index. 
\Remark
Although we are considering the uplink communication in this work, our proposed system model and resource allocation protocol are generic. Therefore, the coverage analysis and joint optimization methodology are well applicable in downlink communication also. Specifically, in downlink  scenario, the total power budget $P_t$ will be associated with the UAV, which it  has to distribute among all the users. For example, UAV is deployed to relieve the stress of terrestrial base station in case of temporary events like festivals and sports and also can be used as flying base stations for public safety scenarios in case of natural disasters. In such cases, the main aim is to serve as many users as possible with available energy resources, which is the objective of this work.  Before moving to problem definition, we do analysis of rate-coverage probability in next section.


\section{Rate-Coverage Probability Analysis} 
In this section, first we  derive the distance distribution of randomly deployed users in the small-cell from the UAV, which is an integral part of further analysis. Then, considering Rician fading, we present closed-form expression for the rate-coverage probability. At last, to get further insight, two special cases of high SNR  and dominant LoS scenario are considered.
\subsection{Distance Distribution of Randomly Deployed Users}
We consider that the users are uniformly distributed  over a circular field of radius $L$ m. With this consideration, the cumulative distribution function (CDF) of random variable $R$ representing the distances of users from the center of circular field can be expressed as $F_R(\tilde{r})=\mathbb{P}(R<\tilde{r})=\frac{\tilde{r}^2}{L^2}$ and its probability distribution function (PDF) can be written as $f_R(\tilde{r})=\frac{2\tilde{r}}{L^2}$, with $0\leq \tilde{r}\leq L$. Now, with the UAV  at an altitude $h$, from the center of circular field, the  distance of a user from the UAV can be expressed as, $\tilde{d}=f(\tilde{r})=\sqrt{\tilde{r}^2+h^2}$, with $\tilde{r}=f^{-1}(\tilde{d})=\sqrt{\tilde{d}^2-h^2}$ and $\frac{\partial \tilde{r}}{\partial \tilde{d}}=\frac{\tilde{d}}{\sqrt{\tilde{d}^2-h^2}}$. With the help of transformation of random variable and denoting $d_{\max}=\sqrt{L^2+h^2}$, we can express  PDF of random variable $D$ representing the distances of nodes from UAV as:
\begin{align}\label{eq:distribution}
f_D(\tilde{d})=f_R(f^{-1}(\tilde{d})){\partial \tilde{r}}/{\partial \tilde{d}}={2\tilde{d}}/{L^2}, \; \forall  \tilde{d}\in\{h,d_{\max}\}. 
\end{align}

Noting this randomness in distances, we combine the distance distribution \eqref{eq:distribution} with Rician fading to continue with our further analysis of rate-coverage probability in next subsection. 

\Remark
The  difference between a traditional 2D network  and the UAV-assisted 3D network is the distance distribution between the serving base station and the users. In the  UAV-assisted network, the third dimension, that is height of UAV from the ground, is also included to calculate the distance between users and UAV, however it is not there in the conventional mobile network. Therefore, the distance distribution in UAV assisted network is over three dimensional space.  Moreover, in the conventional mobile network, Rayleigh fading is commonly used for performance analysis, while in this work for the  UAV assisted network, since there is strong possibility of LoS link, Rician fading is considered for the rate coverage probability analysis. Along with, we have discussed the special case for dominant LoS scenario (Section \ref{sec:HLOS}) which is also highly possible in UAV-assisted communication.
\color{black}
\subsection{Rate-Coverage Probability} 
The rate-coverage probability, $p_{cov_i}$, of a randomly chosen tagged user $U_i$, can be defined as the probability that the assigned data rate to that user is greater than its defined threshold $\eta_{th_i}$, which is expressed mathematically as follows:
\begin{align}\label{eq:COV_P}
p_{cov_i}(\eta_{th_i},P_i,\tau_i)&=\mathbb{E}_{d}[\mathbb{P}[\tau_i\log_2(1+\gamma_i)\geq \eta_{th_i}]]\nonumber\\
&=\mathbb{E}_{d}\left[\mathbb{P} \left[ \gamma_i \geq 2^{ ({\eta_{th_i}}/{\tau_i})}-1 \right] \right]. 
\end{align}
Here $\mathbb{E}_{d}[\cdot]$ is the expectation operator over the random variable $D$  and $\mathbb{P}[\mathcal{E}]$ is representing the probability for happening  event $\mathcal{E}$. The complementary CDF (CCDF) of $\gamma_i$ at $2^{\left( {\eta_{th_i}}/{\tau_i}\right)}-1$, can be further solved as follows:

\begin{align}\label{eq:COV_SNR}
\mathbb{P} \left[ \gamma_i \geq 2^{ {\eta_{th_i}}/{\tau_i}}-1 \right]&=\mathbb{P} \left[ g_i \geq \left(2^{ {\eta_{th_i}}/{\tau_i}}-1\right){\sigma^2}/{P_i} \right]\nonumber\\
&\hspace{-1.2in}\overset{(X_1)}{=}Q_1\left(\sqrt{2K_i}, \sqrt{2(K_i+1)\left(2^{ \frac{\eta_{th_i}}{\tau_i}}-1\right)\frac{{d}^{\alpha}_i\sigma^2}{\mu_i P_i}} \right),
\end{align}
Where, $Q_1(.,.)$ is the first order Marcum Q-function and equality $(X_1)$ comes from the CCDF of non central-$\chi^2$ distribution of channel power gain, 
$g_i$, with mean $\frac{\mu_i}{{d_i}^{\alpha}}$ and Rice factor $K_i$ \cite{grad}. Substituting  \eqref{eq:COV_SNR} in  \eqref{eq:COV_P} and $ b=\sqrt{2(K_i+1)\left(2^{ {\eta_{th_i}}/{\tau_i}}-1\right)\frac{{d}^{\alpha}\sigma^2}{\mu_i P_i}}$, we get:
\begin{align}\label{eq:COV_PF}
p_{cov_i}(\eta_{th_i},P_i,\tau_i)=\int_{h}^{d_{\max}}Q_1\left(\sqrt{2K_i}, b \right)  {2d}/{L^2}\; \text{d}d.
\end{align} 
We removed the subscript $i$ in $d_i$, because we are taking average over distances of all the users from the UAV and here $d$ is treated as a random variable. In next subsection, for analytical insights on the rate-coverage performance over Rician fading channels, we use the approximation for $Q_1(.,.)$ and present the closed-form expression for rate-coverage probability.

\subsection{ Closed-Form Approximation of Rate-Coverage Probability}
 A tight exponential-type approximation \cite{Marcum} for $Q_1(.,.)$ can be expressed as:
\begin{equation}\label{eq:Q_appr}
Q_1(a,b)\approx \exp\left(-e^{\phi(a)}b^{\varphi(a)}\right).
\end{equation}
Here $ \phi(a) $ and $\varphi(a)$ are functions of $a$, and are conditionally defined for lower value of $a$ i.e. $a\ll 1$ and $1\leq a\leq 10$  in \cite{Marcum} and for higher values of $a$ i.e. $10\le a\leq 8000$ in \cite{Ires}. The goodness and reliability of this approximation have been validated in \cite{Ires}, which is good range for our assumption for Rice factors $K_i, \forall i\in I$, and for  dominant LoS  case \ref{sec:HLOS} also in which very high value of Rice factor is considered. 

With this approximation of Marcum Q-function, 
 the tight analytical approximation for $p_{cov_i}$ can be derived as:
\begin{align}\label{eq:rcov1}
p_{cov_i}(\eta_{th_i},P_i,\tau_i)&\approx \int_{h}^{d_{\max}} \exp\left(-e^{\phi\left(\sqrt{2K_i}\right)}b^{\varphi\left(\sqrt{2K_i}\right)}\right)  \frac{2d}{L^2}\; \text{d}d \nonumber\\
&\hspace{-0.35in}\overset{(X_2)}{=}\int_{h}^{d_{\max}} e^{- \mathcal{M}_i d^{\frac{\varphi\left(\sqrt{2K_i}\right)\alpha}{2}}} {2d}/{L^2} \text{d}d\nonumber\\
&\hspace{-0.8in}\overset{(X_3)}{=}\frac{\Upsilon_i\;\mathcal{M}_i^{-\Upsilon_i}}{L^2}\left[\Gamma\left(\Upsilon_i,\mathcal{M}_ih^{\frac{2}{\Upsilon_i}}\right)-\Gamma\left(\Upsilon_i,\mathcal{M}_id_{\max}^{\frac{2}{\Upsilon_i}}\right)\right],
\end{align}
where, equality ($X_2$) is obtained by substituting $\mathcal{M}_i\triangleq e^{\phi(\sqrt{2K_i})}\; \left( 2(K_i+1)\left(2^{ {\eta_{th}}/{\tau_i}}-1\right){\sigma^2}/({\mu_i P_i})\right)^{{\varphi\left(\sqrt{2K_i}\right)}/{2}}$ and equality ($X_3$) comes by using the below identity \cite[2.33.10]{grad} with $m=1$, $n={\varphi\left(\sqrt{2K_i}\right)\alpha}/{2}$, $\delta=\Upsilon_i={4}/({\varphi\left(\sqrt{2K_i}\right)\alpha})$, and $\nu =\mathcal{M}_i$,
\begin{equation}\label{eq:formu}
\int x^m e^{-\nu x^n} dx={-\Gamma(\delta,\nu x^n)}/({n\nu^{\delta}}); \;\delta=({m+1})/{n}.
\end{equation}
\enlargethispage*{2\baselineskip}
 Hence, we can express the rate-coverage probability in closed-form \eqref{eq:rcov1}. We have validated the accuracy of this approximation in Fig$.$~\ref{fig:valid1}, where we can see the good match in both exact and approximate analysis. Although, we have got the closed-form expression of rate-coverage probability, to get further insight, we derive simpler closed-form  expressions for rate-coverage probability  in special cases of high SNR regime, dominant LoS, and Rayleigh-fading scenarios. Since, we have considered Rician fading in channel modeling, dominant LoS and Rayleigh-fading are two extreme cases possible in this channel model. Dominant LoS is very much a feasible scenario due to UAV being at a higher height from the ground with minimum multipath components and similarly non line of sight (NLOS) condition also can occur due to presence of buildings, terrains and other obstacles when UAV is at a lower height.
\subsection{ Rate-Coverage Probability under Some Special Cases}
\subsubsection{High SNR Case}
 In \eqref{eq:Q_appr}, a tight exponential-type approximation for Marcum Q-function is defined as $Q_1(a,b)\approx \exp\left(-e^{\phi(a)}b^{\varphi(a)}\right)$.  Since, under high SNR, $b$ is very low, using the identity $\exp(-x)\approx(1-x)$ for  $x\ll 1$, we can express  $\exp\left(-e^{\phi(a)}b^{\varphi(a)}\right)\approx 1-e^{\phi(a)} b^{\varphi(a)}$, for all  $b\ll 1$. Substituting this in \eqref{eq:rcov1}, we get:
 \begin{align}\label{eq:highSNR}
\widehat{p}_{cov_i}(\eta_{th_i},P_i,\tau_i)&\approx \int_{h}^{d_{\max}}   \frac{2d}{L^2}\left(1-e^{\phi\left(\sqrt{2K_i}\right)}b^{\varphi\left(\sqrt{2K_i}\right)}\right)\; \text{d}d  \nonumber\\
&\hspace{-0.7in}= \frac{d_{\max}^2-h^2}{L^2} - \int_{h}^{d_{\max}} {\frac{2\mathcal{M}_i}{L^2}d^{\frac{\varphi\left(\sqrt{2K_i}\right)\alpha}{2}+1}}\text{d}d \nonumber\\
&\hspace{-0.7in}=1- \frac{2\mathcal{M}_i\left(d_{\max}^{\frac{\varphi\left(\sqrt{2K_i}\right)\alpha}{2}+2} -h^{\frac{\varphi\left(\sqrt{2K_i}\right)\alpha}{2}+2}\right)}{L^2\;\left({\frac{\varphi\left(\sqrt{2K_i}\right)\alpha}{2}+2}\right)}.
 \end{align}

\subsubsection{Dominant LoS (High Rice Factor) Case} \label{sec:HLOS}
Here, we consider that LoS component of received signal is very dominating to its scattered component, and due to that channel has  deterministic constant  power gain $g_i, \forall i\in I$. With this, to derive the rate-coverage probability under dominant LoS scenario, first we define the CCDF of rate assigned to user $U_i,\forall i\in I$ as follows:
\begin{align}\label{eq:rate_CCDF}
\hspace{-0.12in}\mathbb{P}\left[\tau_i\log_2\left(1+\frac{P_i g_i }{d_i^{\alpha}\sigma^2}\right)\geq \eta_{th_i}\right]= \begin{cases}
1;  \text{for}\;\; d_i\leq d_{th_i}\\
0;  \text{otherwise.}
\end{cases}
\end{align}
where $d_{th_i}\triangleq\frac{(P_ig_i)^{\frac{1}{\alpha}}}{(2^{\frac{\eta_{th_i}}{\tau_i}}-1)^{\frac{1}{\alpha}}{(\sigma^2)}^{\frac{1}{\alpha}}}\leq d_{\max}$. Now, we can express the rate-coverage probability as follows:
\begin{align}\label{eq:COV_HR}
{p}^{K}_{cov_i}(\eta_{th_i},P_i,\tau_i)&=\mathbb{E}_{d}\left[\mathbb{P}\left[\tau_i\log_2(1+{P_i g_i }/{(d_i^{\alpha}\sigma^2)})\geq \eta_{th_i}\right]\right]\nonumber\\
&\hspace{-0.5in}\overset{(X_4)}{=}\int_{h}^{d_{th_i}} {2d}/{L^2}\; \text{d}d ={(d_{th_i}^2 - h^2)}/{L^2}\nonumber\\
&\hspace{-0.5in}=\frac{(P_ig_i)^{\frac{2}{\alpha}}L^{-2}}{(2^{{\eta_{th_i}}/{\tau_i}}-1)^{\frac{2}{\alpha}}(\sigma^2)^{\frac{2}{\alpha}}}-\frac{h^2}{L^2}.
\end{align}
Equality $(X_4)$ is obtained by using CCDF of rate given in \eqref{eq:rate_CCDF}. Thus, we  get a simple closed-form expression for rate-coverage probability in dominant LoS scenario.
\subsubsection{Rayleigh Fading Scenario}
Consider that in Rayleigh fading scenario the distance dependent channel power gain $g_i, \forall i\in I$, is exponential distributed with mean $\frac{\mu_i}{d_i^\alpha}$. In this scenario the rate-coverage probability for user $U_i, \forall i\in I$, can be expressed as follows:
\begin{align}\label{eq:COV_PR}
\tilde{p}_{cov_i}(\eta_{th_i},P_i,\tau_i)&=\mathbb{E}_{d}
\left[\mathbb{P}\left[\tau_i\log_2\left(1+{P_i g_i}/{ \sigma^2}\right)\geq \eta_{th_i}\right]\right] \nonumber\\
&\hspace{-0.4in}=\!\int_{h}^{d_{\max}}\!\!\mathbb{P} \!\left[ g_i \geq \left(\!2^{ {\eta_{th_i}}/{\tau_i}}-1\!\right){\sigma^2}/{P_i} \right] {2d}/{L^2}\; \text{d}d\nonumber\\
&\hspace{-0.4in}\overset{(X_5)}{=} \int_{h}^{d_{\max}} \exp\left(-\left(2^{ \frac{\eta_{th_i}}{\tau_i}}-1\right)\frac{{d}^{\alpha}_i \sigma^2}{{\mu}_i P_i}\right) \frac{2d}{L^2}\; \text{d}d\nonumber\\
&\hspace{-0.7in}\overset{(X_6)}{=}\frac{2\;\tilde{\mathcal{M}_i}^{\left(\frac{-2}{\alpha}\right)}}{L^2\alpha}\left[\Gamma\left(\frac{2}{\alpha},\tilde{\mathcal{M}_i}h^{\alpha}\right)-\Gamma\left(\frac{2}{\alpha},\tilde{\mathcal{M}_i}d_{\max}^{\alpha}\right)\right],
\end{align}
where $\tilde{\mathcal{M}_i}=  (2^{ {\eta_{th_i}}/{\tau_i}}-1){\sigma^2}/{({\mu_i} P_i)}$. Equality $(X_5)$ comes using the CCDF of exponential distribution and we get equality $(X_6)$ by using the identity discussed in \eqref{eq:formu}.
In this case: $m=1$, $n=\alpha$, $\delta={(m+1)}/{n}={2}/{\alpha} \; \text{and} \; \nu =\tilde{\mathcal{M}}_i$. 
\begin{Corollary}
The rate coverage probability in \eqref{eq:COV_PR} for Rayleigh fading can be easily formulated by substituting $K_i=0$ (which implies $e^{\phi(\sqrt{2K_i})}=0.5$ and ${\varphi\left(\sqrt{2K_i}\right)}/{2}=1$) in \eqref{eq:rcov1}. With the substitution $K_i=0$ and $\mathcal{M}_i=e^{\phi(\sqrt{2K_i})}\; \left(2(K_i+1)\left(2^{ {\eta_{th}}/{\tau_i}}-1\right) {\sigma^2}/({\mu_i P_i})\right)^{{\varphi\left(\sqrt{2K_i}\right)}/{2}}$  becomes $\tilde{\mathcal{M}_i}$.  
\end {Corollary}

\section{UAV Utility maximization formulation}
In this section, we first formulate an optimization problem  for UAV utility maximization in terms of serving the highest possible number of heterogeneous users under different rate-coverage constraint for each user and energy resources constraints. Then we discuss the challenges in optimally solving  the defined non-convex and combinatorial optimization problem.  At last, we present an iterative feasibility checking  method for fixed integers ranging form lower to upper bound on the number of servable users  to find the global optimal solution for this optimization  problem by utilizing the generalized convexity \cite{Avriel} property of its constraints. 
\enlargethispage*{1\baselineskip}
\subsection{Optimization Problem Formulation}
Focusing on the system model of section II, we consider the problem of allocating the limited power budget $P_t$ and time resources to the heterogeneous users in such a manner that as many users as possible can be served by the UAV, while satisfying their minimum rate requirement $\eta_{th_i}$ bps/Hz and minimum rate-coverage probability requirement $\epsilon_i$ for each user $U_i,\; \forall i\in I\triangleq\{1,2,\ldots,N\}$.  In other words, we want to maximize $N$. Optimal value of integer variable $N$ represents the maximal possible number of user that can be served by the UAV.  So, the proposed design framework is mathematically expressed as:
\begin{align}
(\mathcal{P}):\;&\underset{N,\bold{P},\boldsymbol{\tau}}{\text{maximize}}\quad N\nonumber\\
 \text{s.t.}\; &(C1):   p_{cov_i}(\eta_{th_i},P_i,\tau_i) \geq \epsilon_i,\; \forall i\in I,\nonumber\\
 &(C2): \underset{i=1}{\overset{N}{\sum}}P_i\leq P_t, \quad (C3): \underset{i=1}{\overset{N}{\sum}}\tau_i\leq 1,\nonumber\\ 
  &(C4): N\in \text{Integer}, \quad (C5): P_i\geq 0,\; \forall i\in I,\nonumber\\
  &  (C6): 0 \leq \tau_i\leq 1, \;\forall i\in I,\nonumber
\end{align}
where, constraint $(C1)$ represents the different minimum rate-coverage probability requirement for each user having different rate threshold , $(C2)$ is the power budget constraint,  $(C3)$ is constraint on the time resource budget  which ensures that the sum of normalized time fractions should not be more than one,  $(C4)$ states that number of served users should be an integer, while $(C5)$ and $(C6)$ are the boundary conditions for power and time allocation to each user. The power and time fraction allocation vectors are represented by $\bold{P}=\{ P_1,P_2,...,P_N\}$ and $\boldsymbol{\tau}=\{ \tau_1,\tau_2,...,\tau_N\}$, respectively.
The objective $N$ of this optimization problem $\mathcal{P}$ represents the number of users that can be served by UAV. Therefore, $N$ is an integer variable and along this all the constraints are dependent on $N$. Even the size of vectors $\bold{P}$ and $\boldsymbol{\tau}$ depends on $N$, which itself is unknown. Therefore this problem is a combinatorial, non-convex and NP-hard problem.  Note that, $N$ is acting as both the objective and variable in optimization problem $\mathcal{P}$. It is very much possible for the objective function to be linear function of the variable in an optimization problem, for e.g. ``Linear Programming" \cite{linear_programming} handles the optimization problems having linear objection function in terms of decision variables,  for example $ax+b$, where $a$ and $b$ are constant and $x$ is the optimization variable. In our case $a=1$ and $b=0$. Furthermore, our problem in general belongs to the nonlinear programming class, because although our objective is linear, the constraints are non-linear.

 To solve this problem, we present an iterative feasibility checking method which is discussed in subsection \ref{IFCM}. Before that, in next subsection, we discuss two things that are useful in the iterative feasibility checking method. First, we find the lower  $N_{lb}$ and upper $N_{ub}$ bound on the number of users that can be served by UAV with available power and time resources budget. Second, we discuss the pseudo-concavity property \cite{Avriel} of the rate-coverage probability $P_{cov_i},\; \forall i\in I$,  jointly  in power $P_i$ and time $\tau_i$ allocation variable.

\subsection{Key Insights on Optimal Solution: Bounds and Pseudo-Concavity}\label{lower_upper}
Here, first we define the upper bound $N_{ub}$ on the number of users that can be served by the UAV with available energy resources. 
\subsubsection{ Upper Bound $N_{ub}$ on the Number of Servable Users }
Clearly if the channel condition is favorable and rate-coverage demand is minimum for each user, in this scenario the resources requirement per user will be minimum to fulfill its service demand, i.e., maximum possible $N_{ub}$ users can be served by the UAV with the available energy resources. Let each user has the lowest data rate threshold $\eta_{\min}$, lowest rate-coverage probability $\epsilon_{\min}$ demand, and also the strongest channel condition with maximum mean  channel power gain parameter $\mu_{\max}$. Since, we are considering all the users with equal demand and same channel condition, the resources should be equally distributed among all the users, i.e., $P_i=\frac{P_t}{N_{ub}}$ and $\tau_i=\frac{1}{N_{ub}},\; \forall i\in\{1,...,N_{ub}\}$. Therefore, using \eqref{eq:COV_PF}, $N_{ub}$ can be defined as follows:
\begin{align}\label{eq:N_upper}
N_{ub} \triangleq \left\lbrace N\; \bigg\vert \; \int_{h}^{d_{\max}}Q_1\left(\sqrt{2K}, b_{ub} \right) \frac{2d}{L^2}\; \text{d}d=\epsilon_{\min} \right\rbrace,
\end{align}

 where $b_{ub}=\sqrt{2(K+1)N\left(2^{N \eta_{\min}}-1\right)\frac{d^{\alpha}\sigma^2}{\mu_{\max} P_t}}$. Next, we define the lower bound $N_{lb}$ on the number of users that can be served by the UAV.
\subsubsection{ Lower Bound $N_{lb}$ on the Number of Servable Users}
The users with weakest channel condition and high  rate-coverage demand are bottleneck in UAV utility maximization. If we consider that each user has the highest data rate threshold $\eta_{\max}$, highest rate-coverage probability $\epsilon_{\max}$ demand, and the weakest channel condition with minimum mean  channel power gain parameter $\mu_{\min}$,  the resources requirement per user will be maximum and only $N_{lb}$ users can be served by the UAV. Here also, the resources should be equally distributed among all users, i.e. $P_i=\frac{P_t}{N_{lb}},\,\tau_i=\frac{1}{N_{lb}},  \forall i={1,..., N_{lb}}$. Therefore, using \eqref{eq:COV_PF}, $N_{lb}$ can be defined as follows:
\begin{align}\label{eq:N_lower}
N_{lb} \triangleq \left\lbrace N\; \bigg\vert \; \int_{h}^{d_{\max}}Q_1\left(\sqrt{2K}, b_{lb} \right) \frac{2d}{L^2}\; \text{d}d=\epsilon_{\max} \right\rbrace,
\end{align}
 
where $b_{lb}=\sqrt{2(K+1)N\left(2^{N \eta_{\max}}-1\right)\frac{d^{\alpha}\sigma^2}{\mu_{\min} P_t}}$.

At last, we are proving the generalized convexity property of  rate-coverage probability.
\subsubsection{Pseudo-Concavity Proof for  Rate-Coverage Probability}
Here, we prove the pseudo-concavity property of rate-coverage probability via following lemma.
\begin{Lemma}
$  p_{cov_i}(\eta_{th_i},P_i,\tau_i)$ is jointly pseudo-concave in $P_i$ and $\tau_i$. 
\end{Lemma}  
\begin{IEEEproof}
Using  \eqref{eq:COV_PF}, the rate-coverage probability can be written  as:
\begin{align}\label{PcovQ}
p_{cov_i}=\int_{h}^{d_{\max}}Q_1\left(\sqrt{2K_i}, \sqrt{B(P_i,\tau_i)} \right) {2d}/{L^2}\; \text{d}d,
\end{align}
where, $B(P_i,\tau_i)=\frac{m_i \left(2^{ {\eta_{th_i}}/{\tau_i}}-1\right)}{P_i}$ and $m_i=\frac{2(K_i+1)d^{\alpha}\sigma^2}{\mu_i}$. We express the function $B(P_i,\tau_i)=f_1(P_i)\cdot f_2(\tau_i)$, where, $f_1(P_i)=\frac{m_i}{P_i}$ and $f_2(\tau_i)=(2^{ \frac{\eta_{th_i}}{\tau_i}}-1)$. The first derivative of $f_1(P_i)$ with respect to $P_i$, $\frac{d f_1}{d P_i}=\frac{-m_i}{P_i^2}$ is negative and its second derivative $\frac{d^2 f_1}{d P_i^2}=\frac{2 m_i}{P_i^3}$ is positive. So, $f_1(P_i)$ is  a positive convex decreasing function in $P_i$. Similarly, the function $f_2(\tau_i)$, with $\frac{d f_2}{d \tau_i}=2^{ \frac{\eta_{th_i}}{\tau_i}}\left(\frac{-\eta_{th_i}}{\tau_i^2}\right)$ and $\frac{d^2 f_2}{d \tau_i^2}=2^{ \frac{\eta_{th_i}}{\tau_i}}\left(\frac{\eta_{th_i}^2}{\tau_i^4}+\frac{2 \eta_{th_i}}{\tau_i^3}\right)$ is  a positive convex decreasing function in $\tau_i$. The function $f_1(P_i)$ is independent of $\tau_i$, and $f_2(\tau_i)$ is independent of $P_i$. Therefore, $f_1$ and $f_2$ are jointly positive convex decreasing  in $P_i$ and $\tau_i$. As the product of two positive convex decreasing functions is convex \cite{Avriel}, it infers that  $B(P_i,\tau_i)=f_1(P_i)\cdot f_2(\tau_i)$ is jointly positive convex  in $P_i$ and $\tau_i$. The Marcum Q-function $Q_1\left(\sqrt{2K_i}, \sqrt{B}\right)$ is log-concave and decreasing in $B$ \cite{Logconcave}, and a positive non-increasing log-concave transformation $Q_1(.,.)$ of positive convex function $B$ is log-concave \cite[Lemma 4]{Ires}, i.e., $Q_1\left(\sqrt{2K_i}, \sqrt{B(P_i,\tau_i)}\right)$ is jointly log-concave in $P_i$ and $\tau_i$. Since, log-concavity is preserved under integral operator \cite{boyd}, $   p_{cov_i}(\eta_{th_i},P_i,\tau_i)$ is jointly log-concave in $P_i$ and $\tau_i$. Finally, using \cite[Lemma 5]{Djoint}, we show that a positive differentiable log-concave function $p_{cov_i}(\eta_{th_i},P_i,\tau_i)$ is jointly pseudo-concave in $P_i$ and $\tau_i$.
\end{IEEEproof}

In next subsection, utilizing the pseudo-concavity property of rate-coverage probability and bounds on the number of servable users, we present the iterative feasibility checking method providing global optimal solution of problem $\mathcal{P}$.
\subsection{Iterative Feasibility Checking  Method}\label{IFCM}
We have discussed that the optimization problem $\mathcal{P}$ is combinatorial, non-convex and NP-hard. To  find the global optimal solution of the  optimization  problem $\mathcal{P}$, we present an iterative feasibility checking method. For better understanding of this method, first we focus on the constraints $(C1)-(C6)$. For a fixed integer value of $N$ which is in accordance of the constraint $(C4)$, constraints $(C2)$ and $(C3)$ are the linear constraints. As we have proved in lemma 1, the constraint $(C1)$ is  jointly pseudo-concave in $P_i$ and $\tau_i, \; \forall i\in I$. Thus, with all these properties of constraints, for a given fixed integer value of  $N$, the feasibility of the solution of power and time allocation to serve the $N$ users can be checked by Karush-Kuhn-Tucker (KKT) conditions of $\mathcal{P}$ for fixed $N$ \cite[Th.4.3.8]{Bazaraa}.  The optimization problem  $\mathcal{P}$ for fixed $N$ can be rewritten as:
\begin{align}
(\mathcal{P}_1):\;\underset{\bold{P},\boldsymbol{\tau}}{\text{maximize}}\; N,
 \quad\text{s.t.:}\;   C1,C2,C3,C5,C6.\nonumber
\end{align}

  If $\mathcal{P}_1$ is feasible to serve the  $N$ users by the UAV with available power and time resources, then the optimal $\bold{P}^*$ and $\boldsymbol{\tau}^*$ are given by KKT points the Lagrangian function of $\mathcal{P}_1$. 
     With $\boldsymbol{\lambda}=\{\lambda_1,\lambda_2,..,\lambda_N\}$, $\nu_1$, and $\nu_2$ as Lagrange multipliers for constraints $(C1)$, $(C2)$, and $(C3)$ respectively and keeping the boundary constraints $(C5)$ and $(C6)$ implicit, Lagrangian function of optimization problem $\mathcal{P}_1$ can be written as:
\begin{eqnarray}\label{eq:lagr}
\mathcal{L}(\bold{P},\boldsymbol{\tau},\boldsymbol{\lambda},\nu_1,\nu_2)=N+\underset{i=1}{\overset{N}{\sum}}\lambda_i\left[p_{cov_i}(\eta_{th_i},P_i,\tau_i)-\epsilon_i\right]\nonumber\\
\hspace{-0.2in}-\nu_1\left[\underset{i=1}{\overset{N}{\sum}}P_i-P_T\right]-\nu_2\left[\underset{i=1}{\overset{N}{\sum}}\tau_i- 1\right].
\end{eqnarray}
Apart from   non-negativity of Lagrange multipliers $\boldsymbol{\lambda}$, $\nu_1$, and $\nu_2$,  $3N+2$ KKT conditions are:
\begin{subequations}
\begin{align}\label{eq:lagr_deri}
\frac{\partial\mathcal{L}}{\partial P_i}&=\lambda_i \int_{h}^{d_{\max}}\frac{b_i^2 \bold{\mathrm{I}_o}(a_ib_i)}{2 P_i}\exp\left(-\frac{a_i^2+b_i^2}{2}\right) \frac{2d}{L^2}\; \text{d}d \nonumber \\ & \hspace{1in}-\nu1=0, \;\forall i=1\in I.
\end{align}
\begin{align}\label{eq:lagr_deri1}
\frac{\partial\mathcal{L}}{\partial \tau_i}&=\lambda_i \int_{h}^{d_{\max}}\frac{ \bold{\mathrm{I}_o}(a_ib_i)\eta_{th_i}\ln(2)}{2b_i^{-2}\tau_i^2}\exp\left(-\frac{a_i^2+b_i^2}{2}\right) \frac{2d}{L^2}\; \text{d}d\nonumber\\
& \hspace{1in} -\nu2=0, \; \forall i\in I.
\end{align}
\begin{align}
\lambda_i\left[p_{cov_i}(\eta_{th_i},P_i,\tau_i)-\epsilon_i\right]&=0, \; \forall i\in I,\\
&\hspace{-1in}\nu_1\left[\underset{i=1}{\overset{N}{\sum}}P_i-P_T\right]\!=0,\\
& \hspace{-1in} \nu_2\left[\underset{i=1}{\overset{N}{\sum}}\tau_i- 1\right]\!=0.
\end{align}
\end{subequations}
Here, $\bold{\mathrm{I}_o}(.)$ is the zeroth order modified Bessel function of first kind. Note that there are $3N+2$ variables and $3N+2$ equations, so, the corresponding KKT point can be achieved by solving this system of non-linear equations. If the solved KKT point satisfies the boundary constraints $(C5)$-$(C6)$ and also the non-negativity conditions for Lagrange multipliers, then it will be considered as a feasible solution, i.e.,  $N$  number of users can be served by UAV while satisfying their different data rate and coverage demands. The KKT point defines the optimal power and time allocation for all the $N$ users. Although we are able to check the feasibility of system for a fixed $N$, the main objective of $\mathcal{P}$ is to maximize the number of user served by UAV. So, we have to check the feasibility of KKT conditions of \eqref{eq:lagr} for integer values of $N \in\{N_{lb},N_{ub}\}$, starting from minimum possible $N=N_{lb}$ to the maximum value of $N=N^*$ beyond which KKT equations do not have a feasible solution. This maximum value of $N=N^*$ upto which KKT conditions are feasible, represents the maximum number of users that can be served by UAV by optimally allocating the power and time resources  defined by KKT point. Thus, by using this iterative feasibility check method, the global optimal solution of $\mathcal{P}$ can be achieved.

However, due to the presence of  modified Bessel function  and integration operator in the KKT conditions \eqref{eq:lagr_deri1}  and  
  \eqref{eq:lagr_deri1}, computational complexity to solve this non-linear system consisting of $3N+2$ equations is high. Commercial mathematical packages, for example Matlab, consist of very efficient solvers for non-linear system, but convergence speed of those solvers or conventional sub-gradient methods depends on the starting point and step sizes. Therefore, in the next section, we propose an equivalent transformation solution methodology which can provide the global optimal solution for the optimization problem $\mathcal{P}$ with very low complexity. 

\section{Proposed Solution Methodology}
In this section, first we  propose an energy-efficient equivalent distributed problem formulation in which the centralized utility maximization problem $\mathcal{P}$ is converted into per user energy minimization problem.  Next, we motivate a tight analytical relaxation  for the rate-coverage probability constraint $(C1)$ using Jensen's inequality. Utilizing this relaxation on constraint $(C1)$, we present a variable transformation method that reduces the energy minimization problem with fixed integer $N$ into single dimension problem that can be solved via $N$ parallel computing in distributed manner. We also utilize the variable transformation method for high SNR and dominant LOS scenarios. Then, we provide the low complexity design for joint-optimal solution and present an efficient  joint optimization algorithm to solve the proposed equivalent optimization problem of $\mathcal{P}$. Lastly, we discuss semi-adaptive schemes for solving the proposed equivalent problem 
with individual optimization of  power   and time resource allocation.

\subsection{ Equivalent  Formulation For Per User Energy Minimization }
Main objective of the original problem $\mathcal{P}$ is UAV utility maximization   by optimally allocating   power and time resources to the heterogeneous users. Here, we propose an equivalent transformation of $\mathcal{P}$ to solve it in distributed manner. \begin{Theorem}
The distributed minimization of per user energy consumption, $e_i=P_i \tau_i, \forall i\in I$, while fulfilling their heterogeneous data rate and coverage demand, $ p_{cov_i}(\eta_{th_i},P_i,\tau_i) \geq \epsilon_i,\; \forall i\in I$, is equivalent to  the UAV utility maximization in terms of providing service to the highest  number of users with the limited power $P_t$ Watt and time $T=1$ sec resource budget.
\end{Theorem}
\begin{IEEEproof}
 If we decompose the centralized  problem $\mathcal{P}$ at per user level, it can be solved  in distributed manner via parallel computing. The decomposed problem for user $U_i$ is:
\begin{align}
\hspace{-1in}(\mathcal{P}_0):\;\underset{P_i, \tau_i}{\text{minimize}}\; P_i\tau_i,\nonumber\\
 \text{s.t.:}\;\;   p_{cov_i}(\eta_{th_i},P_i,\tau_i) \geq \epsilon_i; \nonumber\\ \;  0\leq P_i\leq P_{\max}; \; 0\leq \tau_i \leq 1. \nonumber
\end{align}
i.e., if we allocate minimum power $P_i$ and time $\tau_i$ resources to each user $U_i, \forall i\in I$, such that  its data rate and coverage threshold can be achieved, then maximum number of users can be served with available limited power $P_t$ Watt and time resources $T=1$ sec. 
 The minimum possible power $P_i$ and  time $\tau_i$ resource allocation yields minimum energy  consumption $e_i=P_i \tau_i$, in fulfilling the service demand of a users so that maximum number of users can be served by the UAV with the limited resources. Since, we have to distribute the resources among all the users to maximize the number of served users, the sum of energy consumption, $ {{\sum}_{i\in I}}P_i\tau_i$, of all serving users should be minimized.
 
Thus, For a fixed $N$, the energy minimization problem in $\bold{P}$ and $\boldsymbol{\tau}$  can be formulated as:
\begin{align}\label{eqOPT2}
(\mathcal{P}_2):\;\underset{\bold{P},\boldsymbol{\tau}}{\text{minimize}}\quad \overset{N}{\underset{i=1}{\sum}}P_i\tau_i,\quad
 \text{s.t.:}\;   C1,C2,C3,C5,C6.\nonumber
\end{align}

In this formulation  we have to solve the problem $ \mathcal{P}_2$ for different $N$ in range $\{N_{lb}, N_{ub}\}$ to find out the optimal $N^{*}$ i.e. maximum possible $N$ upto which $\mathcal{P}_2$ can be solved with all constraints satisfied. By allocating the resources such that  energy consumed is more than the minimum required energy to fulfill the service demand, the remaining resources will be lesser, and the number of users served will always be less than $N^{*}$. Therefore, we  claim that the unique optimal $N^{*}$ achieved by solving  the energy minimization  problem  $ \mathcal{P}_2 $ for different $N$ in range $\{N_{lb}, N_{ub}\}$ will be equal to that obtained by solving original problem $ \mathcal{P} $. To traverse over the short value space of $N$, we use the golden section search (GSS) method that reduces the search space interval by a factor of 0.618 after each GSS iteration \cite[Ch. 2.5]{BELEGU} and provides fast convergence to global optimal solution $N^{*}$ with optimal resource allocation. In other words, the original non-convex, combinatorial problem  $\mathcal{P}$ is equivalent to solving the optimization problem $\mathcal{P}_2$, $c=\ceil{[(\ln(\Psi)-\ln(N_{ub}-N_{lb}))/(\ln(0.618))]}+1$ times  iteratively for different $N$ defined by GSS method in range $\{N_{lb}, N_{ub}\}$. Here, tolerance  $\Psi\leq 1$, because  variable $N$ is an integer, which is representing the number of users.

Regarding optimal resource allocation there can be two cases: first, if complete resource budget is not used in serving $N^*$ users then the optimal resource allocation  in the original problem $ \mathcal{P} $ can be different from that in  problem $ \mathcal{P}_2 $ with $N=N^*$. However, the remaining resources are not sufficient to serve $(N^*+1)_{th}$ user in both problem formulations. Second, if complete resource budget is used then the optimal resource allocation would be same in both  formulations, because for any resource allocation other than that achieved by solving $ \mathcal{P}_2 $ with $N=N^*$ having minimum possible energy consumption to serve $N^*$ users, optimal $N^*$ can not be achieved.
\end{IEEEproof}

  Next, before defining the joint optimal algorithm to solve $\mathcal{P}_2$, we  motivate an tight analytical relaxation on the probability  $(C1)$ that makes it feasible to transform power variable in terms of time fraction variable and significantly reduces the solution complexity. Please refer Appendix A for low complexity design based on variable transformation approach. 
\subsection{Joint Optimization Algorithm}\label{JOP}
Before defining the joint optimization algorithm, we formulate and solve an optimization problem $\mathcal{P}_3$, which is an integral part of this algorithm.
Using \eqref{eq:Pi_to_ti}, problem $\mathcal{P}_2$ for a fixed integer value of $N$ can be rewritten as:
\begin{align}
(\mathcal{P}_3):\;\underset{\boldsymbol{\tau}}{\text{minimize}}\;\overset{N}{\underset{i=1}{\sum}} V_i\left(2^{ \frac{\eta_{th_i}}{\tau_i}}-1\right)\tau_i \quad \text{s.t.:}\; (C3);  (C6).\nonumber
\end{align}
 Keeping the boundary constraint $(C6)$ implicit and using Lagrangian multiplier $\tilde{\gamma}$ for constraint $(C3)$, the Lagrangian function of optimization problem $\mathcal{P}_3$ for given $N$ can be written as:
\begin{align}
\mathcal{L}_3(\boldsymbol{\tau},\tilde{\gamma})=\underset{i=1}{\overset{N}{\sum}}V_i\left(2^{ \frac{\eta_{th_i}}{\tau_i}}-1\right)\tau_i+\tilde{\gamma}\left[\underset{i=1}{\overset{N}{\sum}}\tau_i- 1\right].
\end{align}
KKT conditions for Lagrange function $\mathcal{L}_3$, can be written as:
\begin{subequations}
\begin{equation}
\frac{\partial\mathcal{L}_3}{\partial \tau_i}=V_i\left(2^{ \frac{\eta_{th_i}}{\tau_i}}-1-2^{ \frac{\eta_{th_i}}{\tau_i}} \frac{ \eta_{th_i}\ln(2)}{\tau_i}\right)+\tilde{\gamma}=0 \quad \forall i\in I
\end{equation}
\begin{equation}\label{eq:comp_slack}
\tilde{\gamma}\left[\underset{i=1}{\overset{N}{\sum}}\tau_i- 1\right]=0. 
\end{equation}
\end{subequations}

\begin{figure}[pt] \centering
\includegraphics[width=9 cm]{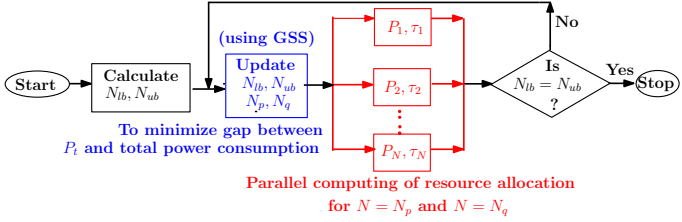}
\caption{Conceptual flow chart of the proposed solution methodology.} \label{fig:flow}
\end{figure}

The complementary slackness condition \eqref{eq:comp_slack} has two possibilities in it. First possibility is $\tilde{\gamma}=0$, it states that ${\sum}_{i=0}^N\tau_i< 1$. We have already made this clear in Section \ref{opt_lambert} that to reduce the energy consumption, maximum possible time should be allocated to the users. So allocating the complete resource of time will tend to global optimal solution i.e minimum possible energy consumption. Therefore, second case will be applicable here, with ${\sum}_{i=0}^N\tau_i-1=0$ and $\tilde{\gamma}$ having some positive value. Now solving $\frac{\partial\mathcal{L}_3}{\partial \tau_i}=0$, which reduces to form $2^x(ax+b)=c$, we get:
\begin{align}\label{Eq:time_sol}
 \tau_i=\frac{\eta_{th_i} \ln(2)}{\bold{W}\left(-(1-{\tilde{\gamma}}/{V_i})2^{({-1}/{\ln(2)})}\right)+1},\;\forall i\in I, 
 \end{align}
which is a closed-form expression in terms of $\tilde{\gamma}$. Note that for high SNR case $V_i=\widehat{V}_i$ and for high Rice-factor case $V_i=V^K_i$.  Substituting $\tau_i, \; \forall i\in I$, in equation $\left[{\sum}_{i=0}^N\tau_i- 1\right]=0$, UAV has to solve just this one equation to get $\tilde{\gamma}$ and it is able to calculate the time $\tau_i$ and power $P_i$ allocated to all $N$ users. Thus, the proposed solution methodology reduces the complexity of solving a system of  $3N+2$ equations  the one solving a single equation for a particular $N$. To find out the optimal $N^{*}$, we propose to iteratively solve $\mathcal{P}_3$ using GSS method upto the maximum $N$ in range $\{ N_{lb}, N_{ub}\}$  for which the constraints $(C2)$ and $(C5)$ can get satisfied. The conceptual flow of the proposed solution methodology is presented in Fig$.$~\ref{fig:flow}.
  
The algorithmic steps for the proposed joint optimization algorithm are summarized in Algorithm 1. Here, all the system and users' parameters namely, power budget $P_t$, noise power $\sigma^2$, radius $L$ of circular region, altitude $h$ of the UAV, path-loss factor $\alpha$, rate threshold $\eta_{th_i}\; \forall i \in I$, coverage threshold $\epsilon_i\; \forall i \in I$, and average channel power gain parameter $\mu_i \; \forall i \in I$ are provided to the Algorithm $1$ as input. Then the optimization problem $\mathcal{P}_3$ is solved for different  values of $N$ in an iterative fashion within the range $\{N_{lb}, N_{ub}\}$ utilizing  GSS method to obtain the optimal number of servable users $N^*$ and joint-optimal power $\bold{P}^*$ and time $\bold{\tau}^*$ resource allocation.
\begin{figure*}[!t]     
	\begin{minipage}{.48\textwidth}       
		\centering\includegraphics[width=3in]{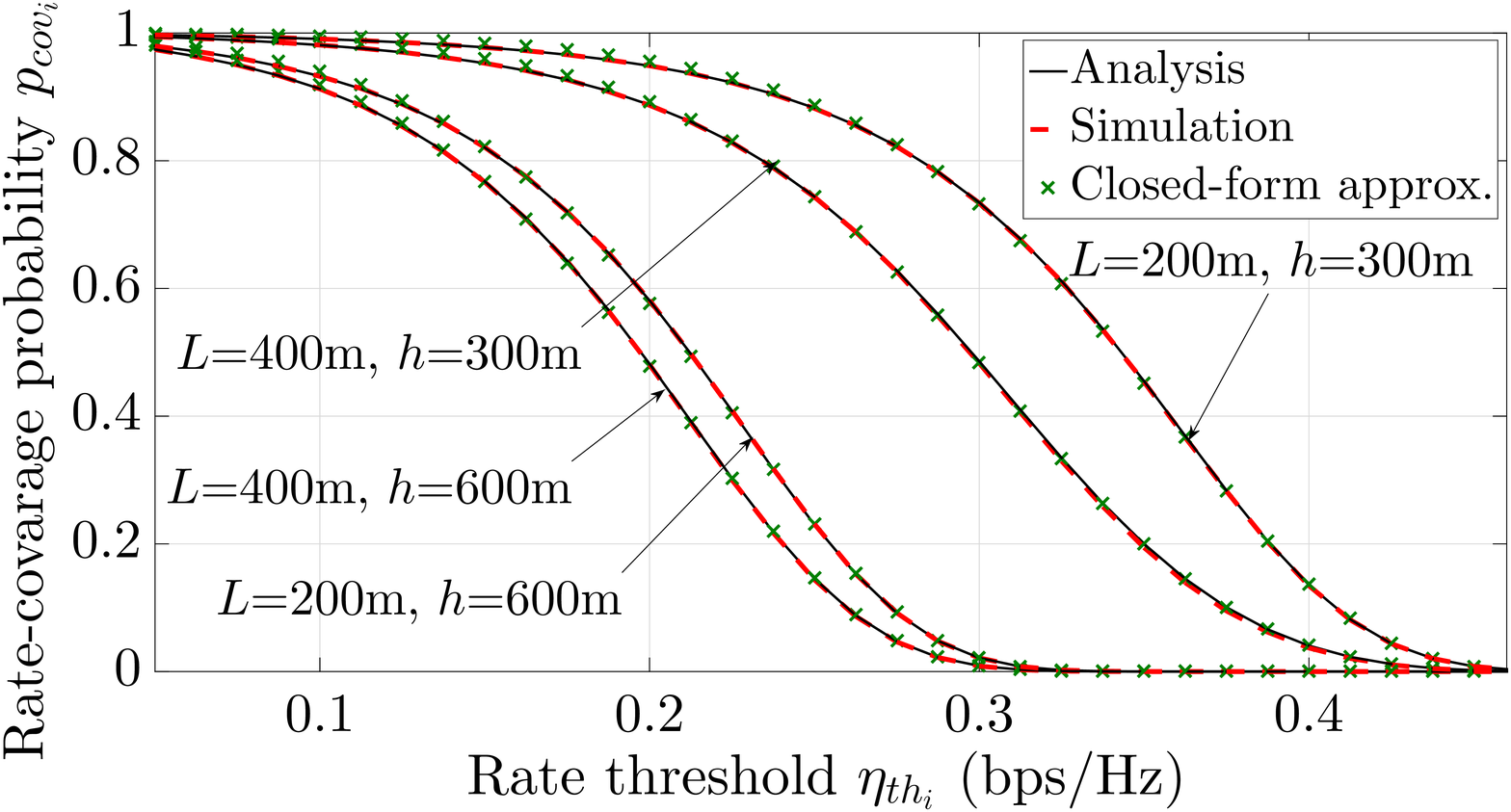}
		\caption{Validation 1: for $p_{cov_i}$ analysis and its closed-form  approximation.}
		\label{fig:valid1}
	\end{minipage}\quad\;  
	\begin{minipage}{.48\textwidth}   
		\centering\includegraphics[width=3in]{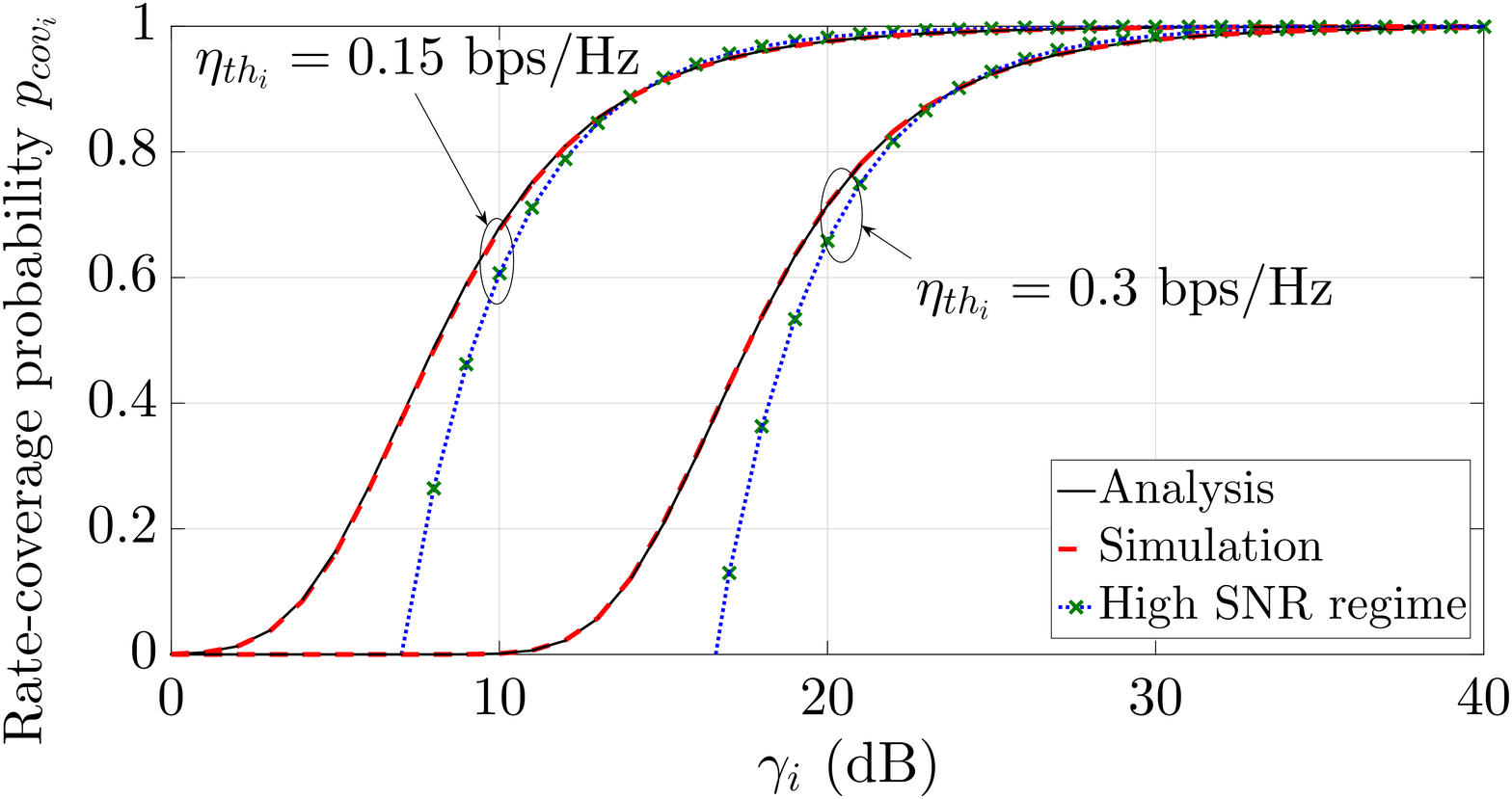}
		\caption{Validation 2: for $p_{cov_i}$ analysis under high SNR regime.}
		\label{fig:valid2}
	\end{minipage}
\end{figure*}

\begin{algorithm}[!h]
\small
{\caption{\small Joint Optimization Algorithm}\label{Algo:ThMax}
\begin{algorithmic}[1]
\Require Channel, system and users' parameters $P_{t}$, $\sigma^2$, $L$, $h$, $\alpha$, $K$, $\eta_{th_i}$, $\epsilon_{i}$, $\mu_i$ $\; \forall i\in I$. 
\Ensure Optimal number of users $N^*$ along with optimal $\bold{P}^*$  and $\boldsymbol{\tau}^*$
\State Obtain $N_{ub}$ and $N_{lb}$ using \eqref{eq:N_upper} and \eqref{eq:N_lower} respectively.
\State Set $N_p=\floor{N_{ub}-0.618(N_{ub}-N_{lb })}$.
\State Set $N_q=\ceil{N_{lb}+0.618(N_{ub}-N_{lb })}$.
\State Solve $\mathcal{P}_3$ with $N=N_p$ and store power and time fraction allocation in $\bold{P_p}$ and $\boldsymbol{\tau_p}$ respectively
\State Solve $\mathcal{P}_3$ with $N=N_q$ and store power and time fraction allocation in $\bold{P_q}$ and $\boldsymbol{\tau_q}$ respectively
\State Set ${\Delta}=N_{ub}-N_{lb}$, and count=0.
\State \textbf{while} ${\Delta}>0$ \textbf{do}
\State \qquad \textbf{if}$\;\;$ $| {P_t-\text{sum}(\bold{P_p})}| \;\leq\; |P_t-\text{sum}(\bold{P_q})|$ $\;\;$  \textbf{then}
\State \quad \qquad Set $N_{ub} = N_q$, $N_q = N_p$, $N_P=\floor{N_{ub}-0.618(N_{ub}-N_{lb})}$.
\State \quad \qquad Set $\bold{P_q}=\bold{P_p}$ and repeat step $4$  to obtain $\bold{P_p}$.
\State \qquad \textbf{else} 
\State \quad \qquad Set  $N_{lb}=N_p$, $N_p=N_q$, $N_q=\ceil{N_{lb}+0.618(N_{ub}-N_{lb})}$.
\State \quad \qquad Set $\bold{P_p}=\bold{P_q}$ and repeat step $5$  to obtain $\bold{P_q}$.
\State \qquad Set ${\Delta}=N_{ub}-N_{lb}$, and count=count+1.
\State Set $N^*=N_{ub}=N_{lb}$.
\State Solve $\mathcal{P}_3$ with $N=N^*$ and store power and time fraction allocation in $\bold{P^*}$ and $\boldsymbol{\tau^*}$
\end{algorithmic}}\vspace{-1mm}
\end{algorithm}

\subsubsection{Convergence Analysis of Algorithm} To find  the unique optimal $N^{*}$, the energy minimization problem $\mathcal{P}_ 3$ is solved for different $N$ in range $\{N_{ lb} , N_{ ub} \}$. To traverse over the short value
space of $N$, we use the GSS method that reduces the search space
interval by a factor of $0.618$ after each GSS iteration [32, Ch. 2.5] and provides fast convergence
to globally-optimal solution $N^{*}$  with optimal resource allocation. In other words,  the optimization problem $ \mathcal{P}_3$ is solved 
$c = [(\ln(\Psi) -\ln(N _{ub} -N_{ lb} ))/(\ln(0.618))] + 1 $ times iteratively for different $N$ defined by GSS
method in range $\{N_{ lb} , N_{ ub} \}$. Here, tolerance $\Psi \leq 1$, because only integer values can be assigned
to variable $N$ , which is representing the number of users. For every iteration in solving $\mathcal{P}_3$, only one equation $(\sum_{i=1}^N \tau_i-1=0)$ has to be solved. Thus, to get optimal $N^*$ and optimal resource allocation using defined algorithm,  the equation $(\sum_{i=1}^N \tau_i-1=0)$ is solved total $c$ times.

\subsection{Individual  Power and Time  Resource Optimization  }
Here, we  discuss the semi-adaptive schemes for energy minimization problem $\mathcal{P}_2$ with only one optimization variable while keeping  other variable fixed.
\subsubsection {Optimal Time Allocation for Fixed Power to All Users}
This case is well suited in the scenario of uplink communication where all users are self dependent for  power source for example each user with  a battery as source of power. Let $P_{\max}$ and $P_{\min}$ denote the maximum and minimum of fixed power available with users, respectively. So,  $\mathcal{P}_2$ can be rewritten as:
\begin{align}
(\mathcal{P}_4):\;\underset{\boldsymbol{\tau}}{\text{minimize}}\qquad \overset{N}{\underset{i=1}{\sum}}P_i\tau_i \quad
 \text{subject to:}\;   C1,C3,C6. \nonumber
\end{align}
By allocating the minimum possible time to each user to fulfill its rate-coverage probability threshold, it will be possible to serve maximum number of users with limited time resource. Using \eqref{eq:Pi_to_ti}, which is the result of satisfying $C1$, we get:
\begin{align}\label{eq:opt_time}
 \tau_i={\eta_{th_i}}/{\log_2(({P_i}/{V_i})+1)}, \; \forall\; i\in I.
 \end{align}
  To find optimal $N^{*}$, $\mathcal{P}_4$ is solved iteratively utilizing GSS method till the maximum $N$ is achieved while satisfying $(C3)$ and $(C6)$.  In this case, upper $N_{ub}$ and lower $N_{lb}$ bounds on $N$ are obtained 
  by substituting $P_i=P_{\max}$ in \eqref{eq:N_upper}, and   $P_i=P_{\min}$ in \eqref{eq:N_lower}, respectively.
   
\subsubsection {Optimal Power Allocation for Fixed Time Allocation to All Users}
Here, we consider the fixed time allocation $\tau_i, \; \forall i\in I$ and want to allocate the limited power resource to all the user optimally such that maximum number of users can be served by the UAV. Let $\tau_{\max}$ and $\tau_{\min}$, respectively, denote the maximum and minimum of fixed time available with users. In this scenario, optimization problem $\mathcal{P}_2$ can be rewritten as:
\begin{align}
(\mathcal{P}_5):\;\underset{\bold{P}}{\text{minimize}}\qquad \overset{N}{\underset{i=1}{\sum}}P_i\tau_i
 \quad \text{subject to:}\;   C1,C2,C5.\nonumber
\end{align}
 The maximum possible number of users can be served with limited power resource if the minimum  power is allocated  to each user to satisfy its rate coverage probability constraint $(C1)$. Thus, using \eqref{eq:Pi_to_ti}, we get:
 \begin{align}\label{eq:opt_power}
  P_i=V_i\left(2^{ {\eta_{th_i}}/{\tau_i}}-1\right), \; \forall\; i\in I.
  \end{align}
   To find optimal $N^{*}$, $ \mathcal{P}_5$ is solved iteratively till the maximum $N$ is reached while satisfying $(C2)$ and $(C5)$, using GSS method.  To reduce the search space for $N$, we obtain  upper $N_{ub}$ and lower $N_{lb}$ bound by  substituting $\tau_i=\tau_{\max}$ in \eqref{eq:N_upper}, and  $\tau_i=\tau_{\min}$ in \eqref{eq:N_lower}, respectively.

   \begin{figure*}[!t]     
	\begin{minipage}{.48\textwidth}       
		\centering\includegraphics[width=3in]{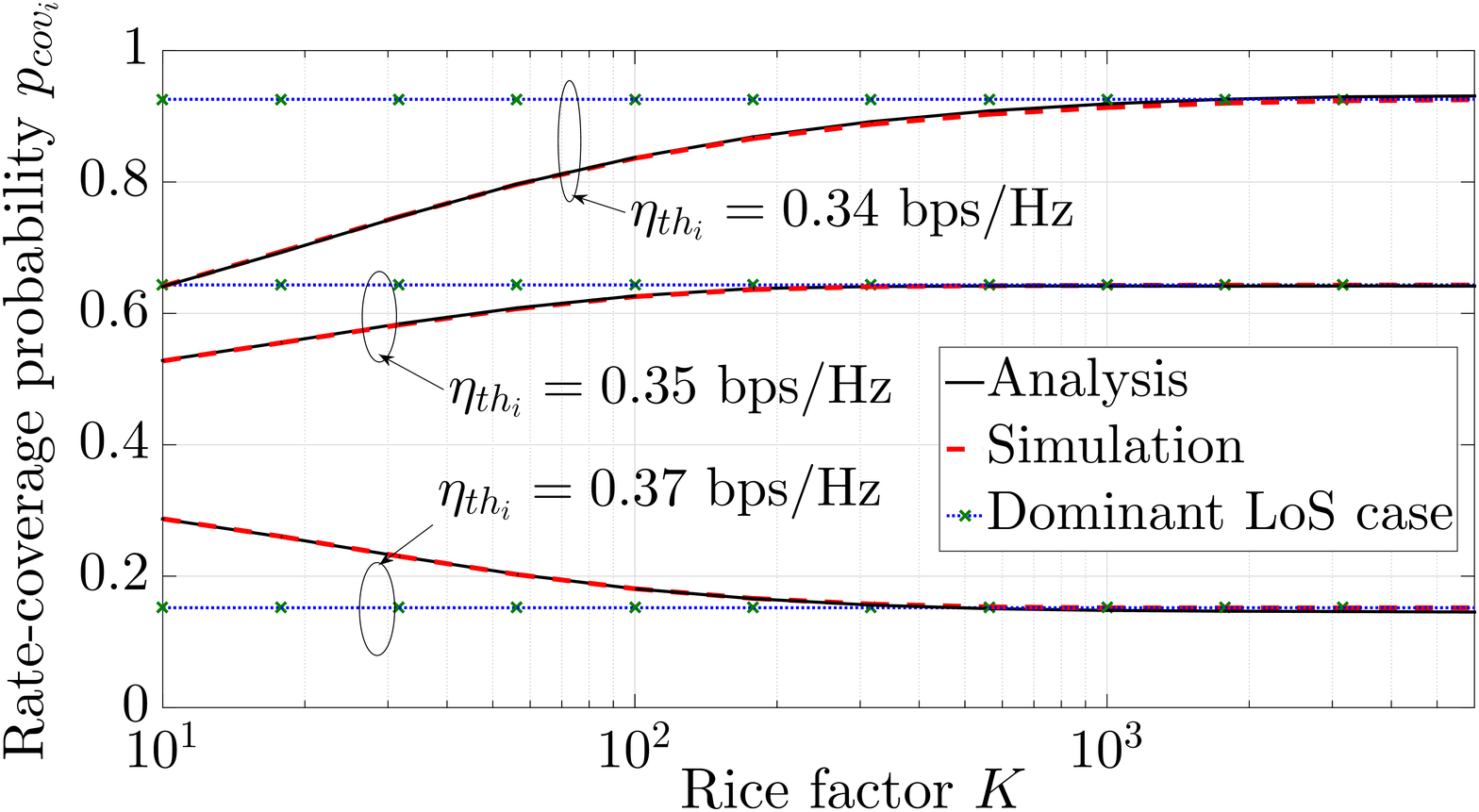}
		\caption{Validation 3: for $p_{cov_i}$ analysis under dominant LoS scenario.}
		\label{fig:valid3}
	\end{minipage}\quad\;  
	\begin{minipage}{.48\textwidth}   
		\centering\includegraphics[width=3in]{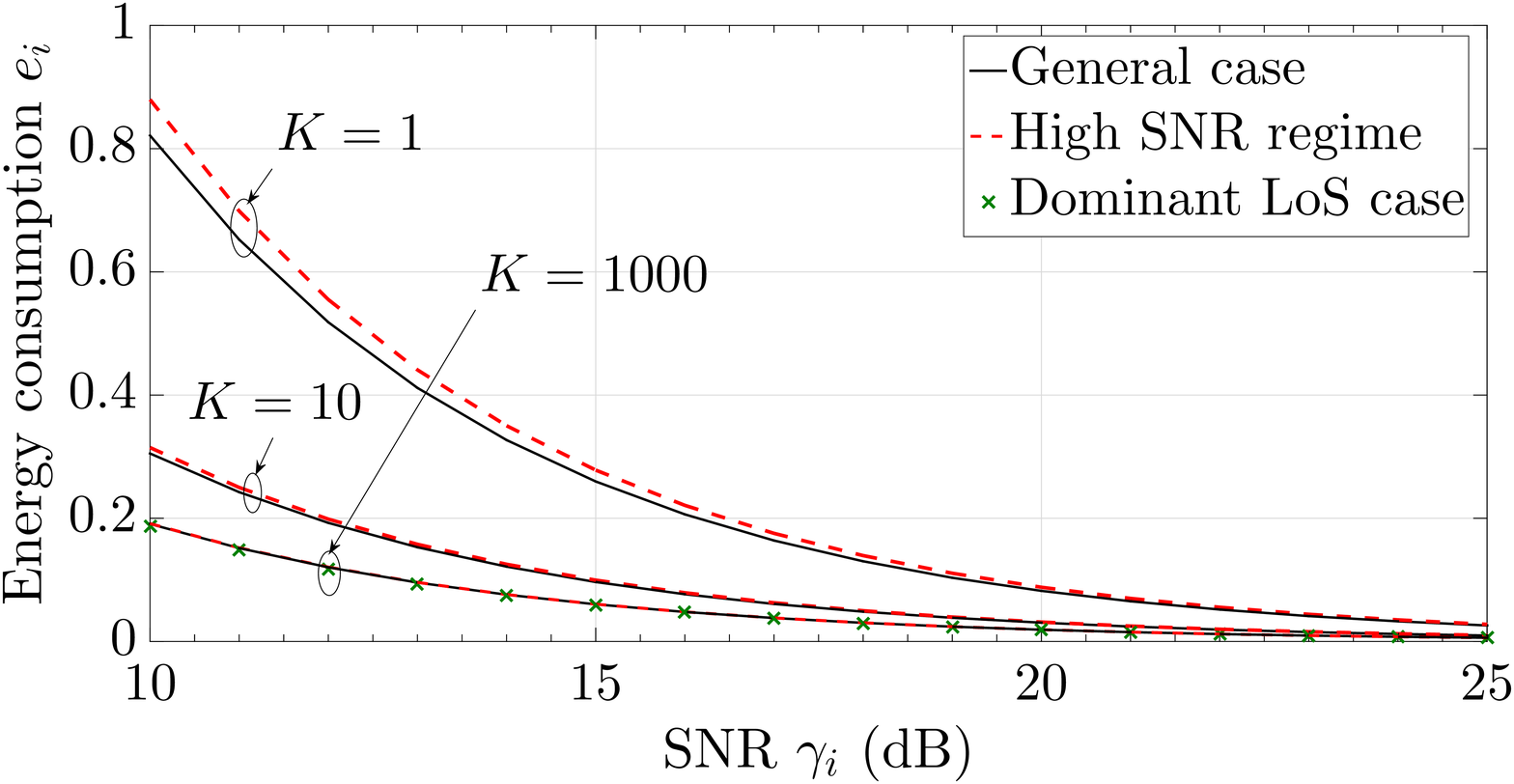}
		\caption{Relative performance of the proposed solution methodology.}
		\label{fig:rel_perform}
	\end{minipage}
\end{figure*}

\section{Numerical Results}
In this section, we first validate the theoretical analysis for the rate-coverage probability and discuss the key design insights on optimal resource allocation. Then, we compare the performance of proposed joint optimization algorithm with the benchmark fixed allocation schemes.   At last, we  evaluate the effect of height of the UAV on its optimal utility while considering the air-to-ground channel modeling, whose analytical details are presented in Appendix B. 

Since, we are considering the users with heterogeneous data rate and coverage demand, to define the service demand for user $U_i$, $\forall i\in I$, we introduce an heterogeneity factor $\beta>0$, base data rate threshold $\eta_{th}$, and maximum coverage demand $\epsilon_{\max}$. With this, the user $U_i$ with index $i$ has data rate requirement $\eta_{th_i}= \eta_{th}\cdot i^{\frac{1}{\beta}}$ and coverage requirement $\epsilon_i= \epsilon_{\max}\cdot \left(\frac{1}{i}\right)^{\frac{1}{a_1\beta}}$. To emulate different channel condition for each user, we consider that average channel power gain parameter for user $U_i$ is $\mu_i=\mu_{th}\cdot i^{\frac{1}{a_2\beta}}$, where $\mu_{th}$ is base average channel power gain parameter. Here, $a_1$ and $a_2$ are constants to limit the  minimum rate-coverage probability and maximum average channel  power gain, respectively. Note that, lower value of $\beta$ corresponds to higher heterogeneity in users' service demand and vice-versa. Unless otherwise stated, the default parameters used in simulation and numerical investigation are: $\alpha=3$, $P_t=30$ dBm, $T=1$ sec, $L=200$ m, $h=400$ m, $K=2$, $\beta=5$, $\mu_{th}=10^{-2}$, $\sigma^2=-90$ dBm, $\eta_{th}=0.1$ bps/Hz, $\epsilon_{\max}=0.99$, and user index $i=1$. Lastly, all the rate-coverage probability results plotted here have been obtained numerically after averaging over $10^5$ independent channel realizations.

\subsection{Validation of Analysis}
First, in Fig$.$~\ref{fig:valid1}, we validate the rate-coverage probability defined in \eqref{eq:COV_PF}, which is in the form of Marcum Q-function, and its closed-form  approximation \eqref{eq:rcov1}  obtained by using the tight exponential approximation of Marcum Q-function. Here, the impact of altitude $(h)$ of UAV, radius $(L)$ of  circular field, and  data rate demand $\eta_{th_i}$  is shown on the rate-coverage probability $p_{cov_i}$ of a tagged user $U_i$. Three main things can be observed from Fig$.$~\ref{fig:valid1}. First, the expressions in  \eqref{eq:COV_PF} and   \eqref{eq:rcov1} have good match with corresponding simulations.  Second, for lower data rate demand $\eta_{th_i}$, $p_{cov_i}$ is higher. Third, with the increase in $h$ or $L$, average distance between users and UAV increases, due to which path loss is higher and corresponding  $p_{cov_i}$ is lower.

\begin{figure*}[!t]     
	\begin{minipage}{.48\textwidth}       
		\centering\includegraphics[width=3in]{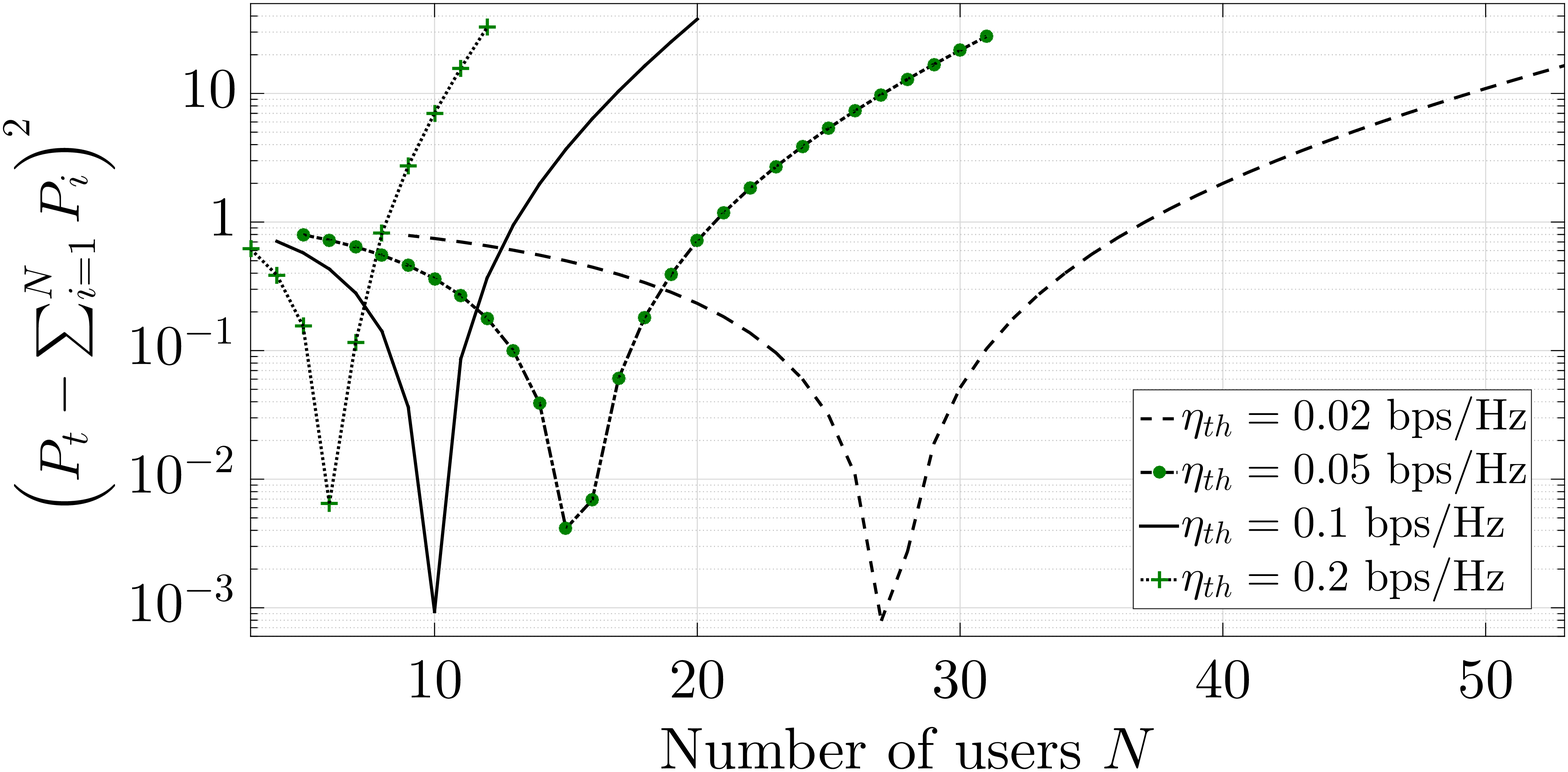}
		\caption{Insight on optimal number of users $N^*$ along with the verification of the tightness of bounds and unimodality claim.}
		\label{fig:opt_N}
	\end{minipage}\quad\;  
	\begin{minipage}{.48\textwidth}   
		\centering\includegraphics[width=3in]{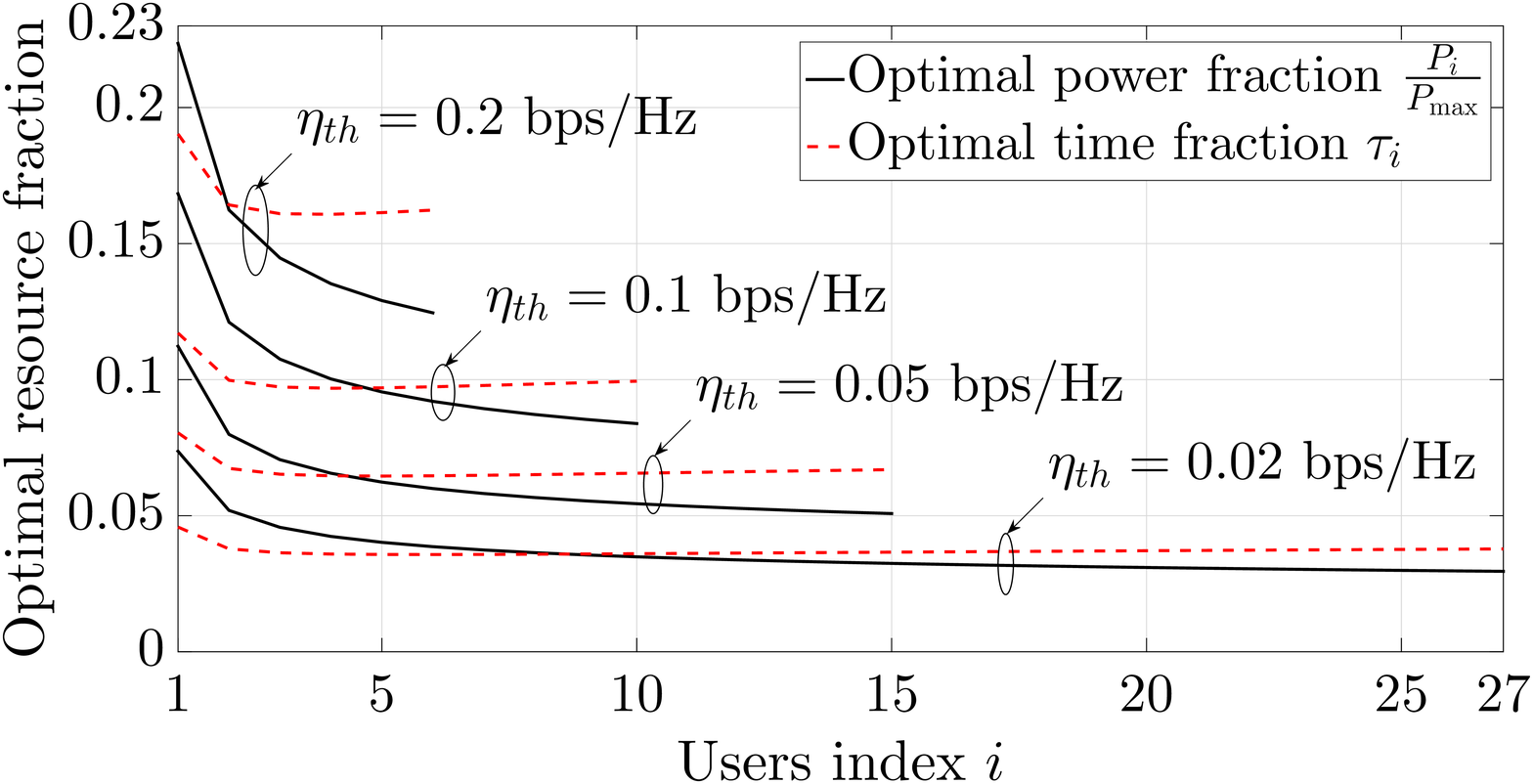}
		\caption{Insight on optimal time and power allocation with optimal  $N^*$ for different base rate threshold $\eta_{th}$.}
		\label{fig:opt_power_time}
	\end{minipage}
\end{figure*}
Next, in Fig$.$~\ref{fig:valid2} and Fig$.$~\ref{fig:valid3}, we validate the rate-coverage probability analysis for special cases of high SNR \eqref{eq:highSNR} and dominate LoS scenario \eqref{eq:COV_HR}, respectively. In Fig$.$~\ref{fig:valid2}, it can be observed that equation \eqref{eq:highSNR} matches well with the simulation and analysis (using Marcum Q-function \eqref{eq:COV_PF}) in high SNR regime. One point can be noted from this figure is that for lower $\eta_{th_i}$, the SNR value where equation \eqref{eq:highSNR} have good match with  analysis and simulation is lower. 
Fig$.$~\ref{fig:valid3} validates the rate-coverage probability analysis for dominant LoS scenario, i.e., high Rice factor K case \eqref{eq:COV_HR}. It can be observed from this figure that for high $K$ values, equation  \eqref{eq:COV_HR} has close match with simulation and analysis \eqref{eq:COV_PF}. Since we have validated the analysis of rate-coverage probability through simulation, we  utilize the analytical results in further investigations.

\subsection{Key insights on optimal resource allocation to users}
 In this subsection, in Fig$.$~\ref{fig:rel_perform}, we check the relative performance of the proposed solution methodology presented in Section V for the following three scenarios:  first, for the general case where the tight analytical approximation is utilized on the  constraint $(C1)$, second, for high SNR regime, and third, for dominant LoS scenario.  The main use of the special cases such as high SNR and dominant LoS is to prove the reliability of our solution methodology because in these special cases exact closed-form expression of rate coverage probability is used without any relaxation in constraint $C1$. It has been proved in Theorem 1 that maximization of number of user under service is equivalent to minimization of energy consumption per user to fulfill its rate and coverage requirement. Therefore, we are comparing the relative performance in terms of energy consumption $e_i$. In this figure, it can be observed that the performance of proposed methodology for high SNR regime is approaching to that in general case for high SNR $\gamma_i$ values.  For lower SNR $\gamma_i$ values, the proposed methodology in general case is providing lower energy consumption to fulfill user's demand. For high Rice factor $K$ values,  energy consumption by proposed methodology in dominant LoS scenario  is almost equal to that in general case.  The  solutions for special cases of high SNR regime  and dominant LoS scenario have limitation to be applicable only for high SNR and  high $K$ values, respectively, while the proposed methodology for general case is valid in all scenarios.

Next, Fig$.$~\ref{fig:opt_N} gives insight on the optimal number on users $N^*$ that can be served with the available energy resources while full-filling the heterogeneous demand of all the $N^*$ users. It is proved in Appendix \ref{low_complexity} that to minimize the energy consumption, which in turn will maximize the number of users under service, complete block time should be allocated to users optimally. Therefore, optimal number of users is the highest possible number of users that can be served within  power budget constraint. That is why on y-axis of Fig$.$~\ref{fig:opt_N} we have shown the squared difference between the available power budget $P_{t}$ and sum of optimal power used in serving $N$ users. { With the increase in $N$, the sum $\sum_{i=1}^{N}P_i$ also increases. The  value of $N$ for which the difference $P_t- \sum_{i=1}^{N}P_i$ is minimum and positive,  that $N$ will be the optimal number of users $N^*$ servable with available energy resources. If $i= (N^*+1)_{th}$ user is served by the UAV, the difference $P_t- \sum_{i=1}^{N^*+1}P_i$ would be negative.  From Fig$.$~\ref{fig:opt_N}, it can be observed that for a fixed rate threshold $\eta_{th}$, optimal $N^*$ is unique and therefore, the objective function of utility maximization problem is unimodal in $N$.}  We have plotted the squared power difference for different values of base rate thresholds $\eta_{th}$. Plots are drawn from lower $N_{lb}$  to upper $N_{ub}$ bounds on the number of users for different base rate thresholds $\eta_{th}$. With higher rate threshold $\eta_{th}$, lesser number of users can be served and also the gap between lower and upper bound is small. Therefore, the solution complexity to obtain optimal $N^*$ is lesser in case of higher  base rate thresholds $\eta_{th}$ and vice versa.

 Fig$.$~\ref{fig:opt_power_time} represents the optimal fraction of power and time resources allocated to optimal number of users $N^*$. This global optimal resource allocation solution is  outcome of the proposed joint optimization algorithm presented in Algorithm 1. We have considered the same four base rate thresholds $\eta_{th}$  as in Fig$.$~\ref{fig:opt_N}. For example, for $\eta_{th}=0.1$ bps/Hz the optimal number of users are $N^*=10$. In Fig$.$~\ref{fig:opt_power_time}, it is shown that how total power and time resources are allocated to $N^*$ users for specific base rate threshold $\eta_{th}$.  For higher data rate threshold $\eta_{th}$, more resources $(P_i,\tau_i)$ are allocated to full-fill their demand. That is why with fixed resources, lesser number of users can be served when they have higher data rate demands.  For user index $i=1$, $\epsilon_i$ is highest, $\epsilon_i=\epsilon_{\max}=0.99$, this is the reason that to fulfill its coverage demand highest fraction of power and time resources are allocated to it. Trend for time allocation is such that total block time is almost equally distributed among all the $N^*$ users. Power allocation is slightly decreasing with the increase in user index number, reason behind this is that $\epsilon_i= \epsilon_{\max}\cdot (1/i)^{{1}/{a_1\beta}}$ is decreasing and average channel gain parameter $\mu_i=\mu_{th}\cdot i^{({1}/{a_2\beta})}$ is increasing with the increase in  user index $i$.
 
\begin{figure*}[!t]     
	\begin{minipage}{.48\textwidth}       
		\centering\includegraphics[width=3in]{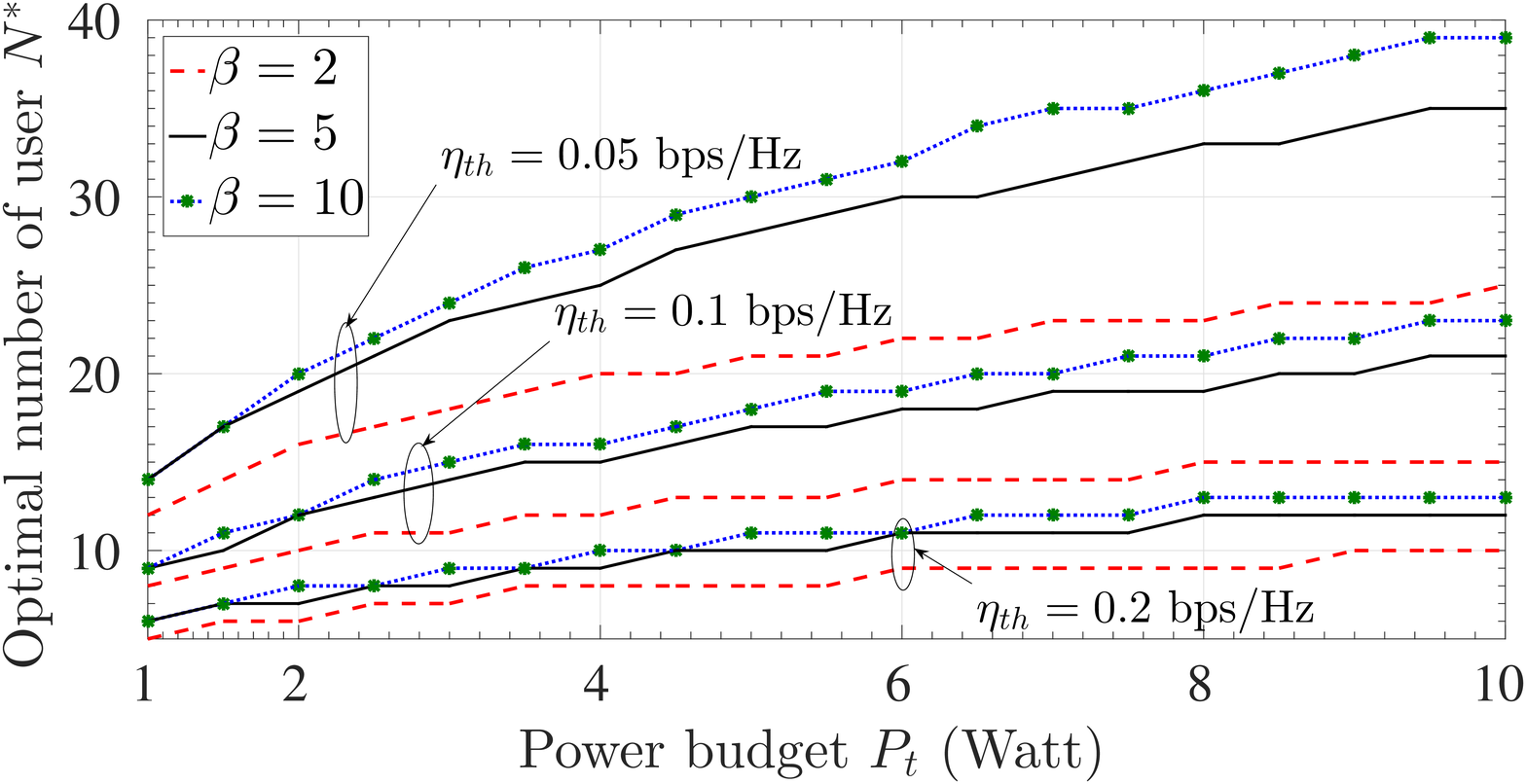}
		\caption{Optimal number of users $N^*$ with different power budget $P_t$, heterogeneity factor $\beta$, and base rate threshold 
			$\eta_{th}$.}
		\label{fig:opt_N_PMAX}
	\end{minipage}\quad\;  
	\begin{minipage}{.48\textwidth}   
		\centering\includegraphics[width=3in]{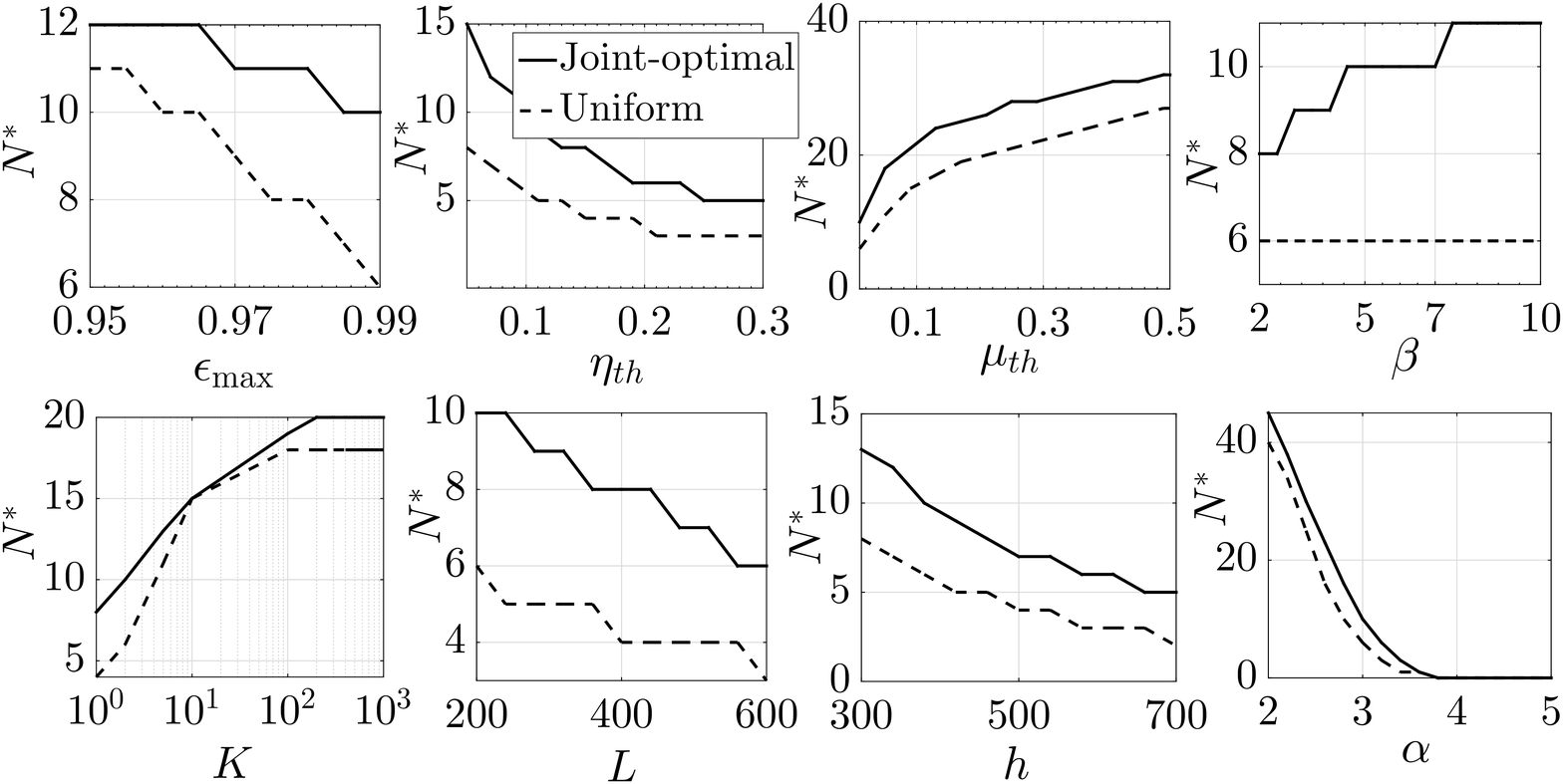}
		\caption{Optimal number of users $N^*$ for different parameters using joint-optimal and uniform resource allocation scheme.}
		\label{fig:opt_N_all}
	\end{minipage}
\end{figure*}

\subsection{Design Insight and Performance Comparison With Benchmark Schemes}
Here, first we represent the effect of different system and users' parameters on the optimal number of users $N^*$. Then, the performance comparison of our proposed (joint-optimal) approach is done against the fixed benchmark schemes having  the uniform and semi-adaptive resource allocation. The reason behind the selection of  the defined benchmark schemes for the comparison purpose is that the related works have very different objectives while proposing the resource allocation technique, such as,  packet transmission delay minimization \cite{JLI}, maximizing the throughput \cite{RFAN, Wang_joint}, and transmission power consumption minimization \cite{Soorki_M2M}. The comparison of the proposed scheme with the schemes in literature is not fair due to having different objectives.   

Fig$.$~\ref{fig:opt_N_PMAX} represents the effect of total power budget $P_t$, base rate threshold $\eta_{th}$, and the heterogeneity factor $\beta$ on the optimal number of users $N^*$. Three points to be noted in this figure are: first, with the increase in $P_t$, the optimal $N^*$ increases, i.e., if power resource is more, rate and coverage requirement can be full-filled for larger number of user, and hence, optimal $N^*$ increases with increase in power budget. Second, with the decease in base rate threshold $\eta_{th}$, the optimal $N^*$ increase. The reason behind this is that to full-fill lower data rate requirements, lesser resources are needed and the remaining resources can be used to serve extra users, that is why with same energy resource budget, the optimal $N^*$ increases with decrease in base rate threshold $\eta_{th}$. Third, with the increase in heterogeneity   factor $\beta$, the optimal $N^*$ increases, because if $\beta$ is low, users demand would be more heterogeneous and highly increasing with the user index number, and lesser number of user would be able to get served with available resource. Whereas for high value of $\beta$, users' data rate demand are almost same as the base rate threshold and higher number of users can be served with available energy resources.  

In Fig$.$~\ref{fig:opt_N_all} effects of different system parameters are shown on the optimal number of users $N^*$ for fixed available energy resources. Here, solid  and dash lines represent the optimal $N^*$ for joint-optimal and  uniform/equal resource allocation schemes, respectively. If there is increment in rate-coverage probability $\epsilon_{\max}$ demand or in base rate threshold $\eta_{th}$ requirement, it need more resources to fulfill the user's demand, and lesser number of user would be able to get required service with available energy resources. With higher base average channel gain parameter $\mu_{th}$, channel condition will be better and lesser resources would be required to serve a user, in effect of which larger number of users can get service with available energy resources. With increase in heterogeneity factor $\beta$, there is less increment in users' rate demand with the user index number or in other words larger number of users can get their requirement full-filled in comparison the case with lower $\beta$, i.e., high heterogeneity in users' demand where users demands highly increase with increase in user index number.  It can be observed that for higher value of $B$, larger utility gain is achieved by the proposed scheme over uniform resource allocation scheme.  Rice factor $K$ is directly related to mean channel power gain in LoS component, therefore with increase in $K$, users receive higher power from LoS component even for fixed transmission power. Hence, with higher $K$, lesser transmission power is need to be allocated to fulfill the user rate and coverage requirement, in other words with same available energy resource optimal $N^*$ is higher for higher Rice-factor $K$. The utility gain is more for the lower and upper range of $K$, while in its  middle range utility gain is less. With  the increase in sparsity in users over the circular field for its larger radius $L$ or with the increase in UAV altitude $h$, the distance between UAV and users will increase, which in turn will increase the path loss. The increase in path loss factor $\alpha$ due to lossy environment also increases the path loss. To overcome this increment in path loss, more resources are needed to provide desired service to users. Hence, with the increase in $L$, $h$, and $\alpha$ lesser number of users would be able to get service with available energy resources.

\begin{figure*}[!t]     
	\begin{minipage}{.48\textwidth}       
		\centering\includegraphics[width=3in]{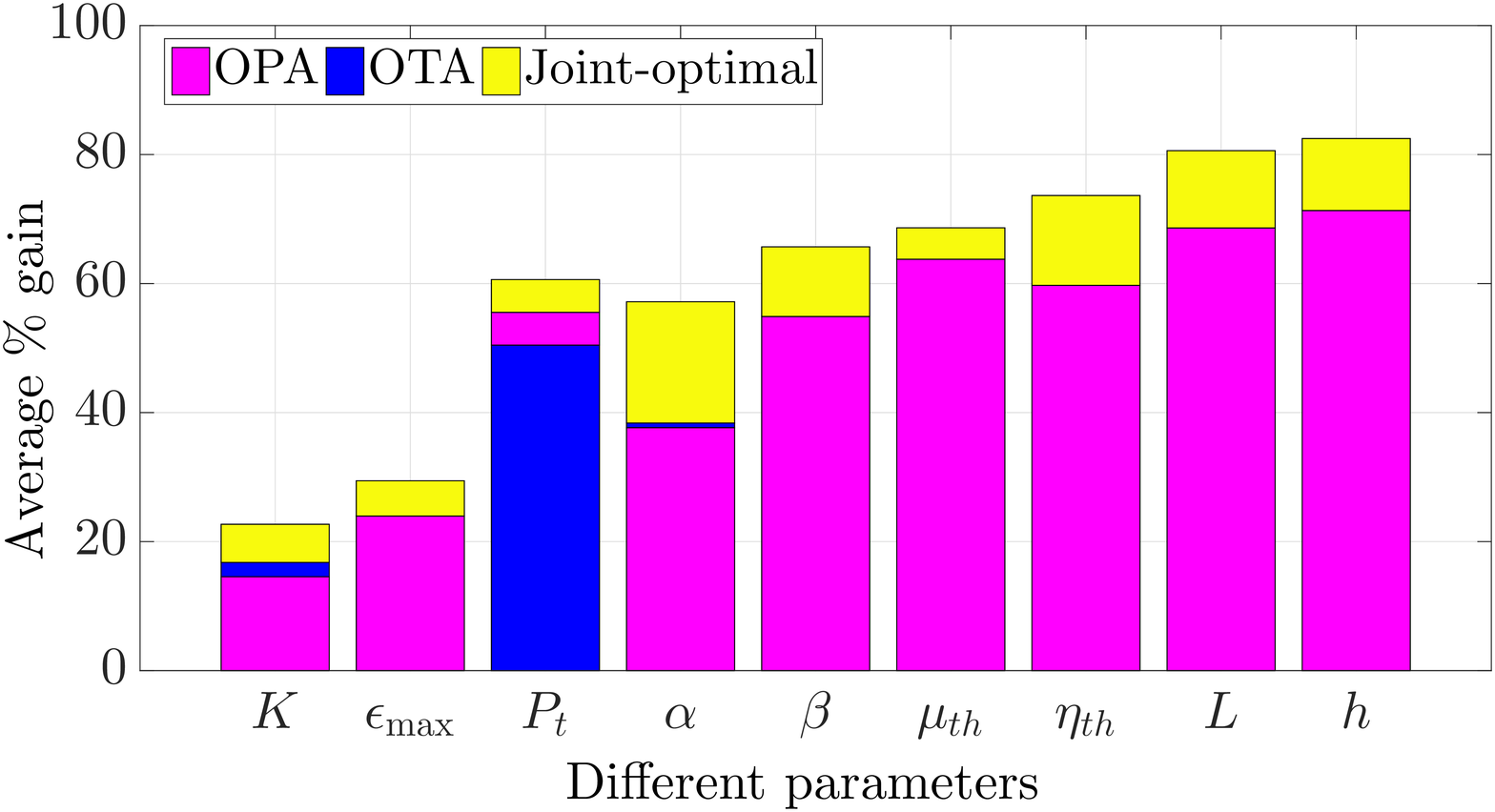}
		\caption{Percentage gain in optimal number of users achieved by proposed schemes against uniform resource allocation scheme.}
		\label{fig:gain}
	\end{minipage}\quad\;  
	\begin{minipage}{.48\textwidth}   
		\centering\includegraphics[width=3in]{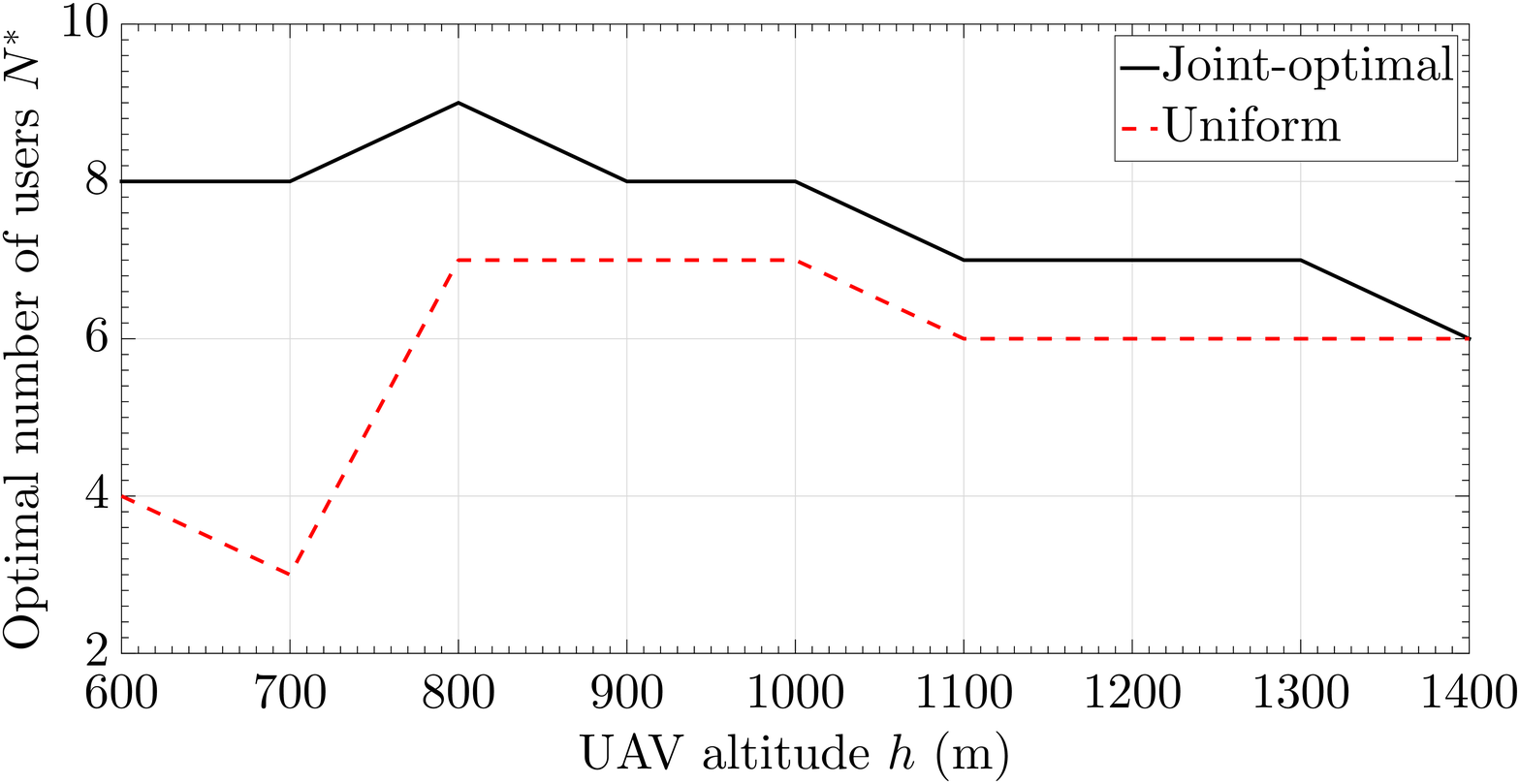}
		\caption{Performance comparison of proposed joint-optimal algorithm  with benchmark schemes considering air-to-ground channel modeling.}
		\label{fig:perform_LOS_NLOS}
	\end{minipage}\vspace{-4mm}
\end{figure*}
At last, Fig$.$~\ref{fig:gain} compares the performance of individual optimization schemes (OPA: optimal power allocation having uniform time allocation; OTA: optimal time allocation having uniform power allocation) and joint-optimal schemes (joint-optimal power and time allocation) discussed in Algorithm 1, against the uniform allocation scheme (uniform power and uniform time allocation). This figures represents the  average percentage gain of OPA, OTA, and joint-optimal schemes over the uniform allocation scheme in terms of UAV Utility. We have considered nine different parameters on x-axis, and y-axis is representing the marginal average percentage $(\%)$ gain. To calculate  average $\%$ UAV utility gain for a parameter, first we analyzed the $\%$ gain in optimal $N^*$ for different value of parameter in the range shown in  Fig$.$~\ref{fig:opt_N_PMAX} and Fig$.$~\ref{fig:opt_N_all}, and then averaged it. Three points to note in this figure are:  first, we observe that for  Rice factor $K$ and path-loss factor $\alpha$, OTA scheme is providing higher UAV utility gain in comparison of OPA. Reason behind this is that in the scenarios, where rice factor $K$ and path loss factor $\alpha$ vary in ranges $(1, 1000), (2, 5)$,
respectively, signal-to-noise-ratio (SNR) is more controlled by these factors  than the power
allocation. That is why in assigning required rate to a user, optimal time fraction allocation has more effect than the optimal power allocation. Thus, OPA is providing less utility gain for these cases in comparison of OTA. Second, with variation in power budget $P_t$, OPA scheme is better than OTA because if power budget $P_t$ is considered to vary from $1$ to $10$ Watt while keeping all other parameters fixed, allocating power optimally using proposed OPA according to the requirement of users can result in serving higher number of users in comparison of the OTA, where power is equally distributed among all the users. Since, OPA is more sensitive to $P_t$ variation than OTA, OPA has higher utility gain than that in OTA  by varying the power budget $P_t$.  For remaining parameters $\%$ UAV utility gain for both OPA and OTA schemes are equal over the uniform allocation scheme and that is why with marginal gain representation only one bar color is visible for these parameters. In comparison of benchmark uniform resource allocation scheme, the proposed  OPA and OTA schemes provide  $49.56\%$ and $49.77\%$ gain, respectively in the UAV utility. Third, joint-optimal resource allocation scheme is performing better than OPA and OTA for all nine parameters and provides maximum $82\%$  UAV utility gain for parameter $h$ (UAV altitude) and minimum $23\%$ UAV utility gain for parameter $K$ (Rice factor) and on an average $59.66\%$ UAV utility gain for all parameters over the uniform allocation scheme. In other words, by using the proposed joint-optimal resource allocation scheme, almost $60 \%$ more users can be served with the available energy resources in comparison of uniform resource allocation scheme.  

\subsection{ Evaluation of Effect of Height of UAV  with Air-to-Ground Channel Consideration}
Fig. \ref{fig:perform_LOS_NLOS} presents the comparison of proposed joint-optimal  algorithm in terms of optimal number of servable users $N^*$ with the benchmark uniform resource allocation scheme  with the consideration of air-to-ground channel modeling. The  parameters used in simulation and numerical investigation in this case are: $\alpha=3$, $P_t=21$ dBm, $T=1$ sec, $L=500$ m, $\beta=2$, $\sigma^2=-90$ dBm, $\eta_{th}=0.2$ bps/Hz, $\epsilon_{\max}=0.99$, and $B=0.136$ and $C=11.95$ (for dense urban scenario). Since the LoS link probability increases with the height of the UAV and also the path loss increases with the increase in distance between UAV and  users, there is an optimal height at which utility of the UAV can be maximized by serving the highest number of users. It can be seen from Fig. \ref{fig:perform_LOS_NLOS} that for the given parameters the optimal height of UAV is $800$ m. It also can be  observed in this figure that the proposed solution methodology perform better than the uniform resource allocation scheme with air-to-ground channel modeling. Thus, our proposed scheme is well applicable with air-to-ground channel modeling also.

\section{Conclusion}
This paper has investigated energy-efficient joint-optimal power and time resource allocation to heterogeneous users in UAV-assisted communication network in order to maximize the UAV utility by serving the highest possible number of users  with available energy resources while fulfilling their different data rate and coverage requirement. Closed-form expression for rate-coverage probability over Rician fading channels has been derived and  extended the analysis also for special cases of high SNR regime and dominant LoS scenario  providing simpler closed-form expression for rate-coverage probability to get further analytical insight. By proving the generalized convexity  of the optimization problem for fixed $N$, an iterative feasibility checking method has been proposed having  search space defined by lower and upper bounds on the number of servable users with available energy resources. To further reduce the solution complexity, an equivalent distributed energy minimization problem has been solved via parallel processing by utilizing a tight analytical relaxation on rate-coverage probability constraint and variable transformation method, which reduce it into single variable problem providing closed-form expression for joint-optimal power and time allocation solution. The steps used in proposed solution methodology have been well listed in joint-optimization algorithm. It has been observed that to maximize the UAV utility, time resources should be completely distributed among the users under service. Individual optimization schemes  of power and time resource allocation also has been investigated. The analysis is validated by simulation results. Via numerical investigation, design insights using different system and users' parameters have been discussed for the optimal number of users and for the optimal resource allocation. In comparison of benchmark  uniform resource allocation scheme, the proposed  OPA, OTA and Joint-optimal schemes provide  $49.56\%$, $49.77\%$, and $59.66\%$ gain, respectively, in the UAV utility.
\appendices
\section{Low Complexity Design Based on Variable Transformation Approach} 
In this section,  first we propose a tight analytical relaxation on the rate-coverage probability constraint $(C1)$ for the general case \eqref{eq:rcov1}  and transform the power variable $P_i$ in terms of time fraction variable $\tau_i$. Then, we consider rate-coverage probability constraint $(C1)$ for both special cases of high SNR regime \eqref{eq:highSNR} and dominant LoS scenario \eqref{eq:COV_HR} and perform the variable transformation.  At last, using this variable transformation approach, we propose a low complexity joint optimization design for per user energy minimization.

\subsubsection{Proposed Analytical Relaxation and Variable transformation For General Case}
Using \eqref{eq:rcov1}, the constraint $(C1)$ can be written as $\int_{h}^{d_{\max}} \exp\left(-e^{\phi\left(\sqrt{2K_i}\right)}b^{\varphi\left(\sqrt{2K_i}\right)}\right) \frac{2d}{L^2} \text{d}d \geq \epsilon_i$, where $b=\sqrt{2(K_i+1)\left(2^{ {\eta_{th_i}}/{\tau_i}}-1\right){({d}^{\alpha}\sigma^2)}/{(\mu_i P_i)}}$ includes  both optimization variables $P_i$ and $\tau_i$ in it. 
In this form of $(C1)$, the  closed-form representation of one optimization variable in terms of other  variable is not possible.  To do this, by taking $\log$ on both sides of $(C1)$  and we get
\begin{align}
\tilde{C}=\log\left( \mathbb{E}_{d}\left[\exp\left(-e^{\phi\left(\sqrt{2K_i}\right)}b^{\varphi\left(\sqrt{2K_i}\right)}\right)\right ]\right) \geq \log(\epsilon_i)=\tilde{B}.
\end{align}
By following the Jensen's inequality, we can write $\tilde{C}=\log\left( \mathbb{E}_{d}\left[\exp\left(-e^{\phi\left(\sqrt{2K_i}\right)}b^{\varphi\left(\sqrt{2K_i}\right)}\right)\right ]\right)\geq \mathbb{E}_{d}\left[\log\left(\exp\left(-e^{\phi\left(\sqrt{2K_i}\right)}b^{\varphi\left(\sqrt{2K_i}\right)}\right)\right)\right ]=\tilde{A}$. If $\tilde{C}\geq \tilde{A}, \tilde{A}\geq \tilde{B}$, we get $\tilde{C} \geq \tilde{B}$, With this constraint $C1$ approximates to 
\begin{align}
 \mathbb{E}_{d}\left[-e^{\phi\left(\sqrt{2K_i}\right)}b^{\varphi\left(\sqrt{2K_i}\right)}\right ] \geq \log(\epsilon_i).
\end{align}
This can be expressed  as:
\begin{align}
-  \left( 2(K_i+1)\left(2^{ \frac{\eta_{th_i}}{\tau_i}}-1\right)\frac{\sigma^2}{\mu P_i}\right)^{\frac{\varphi\left(\sqrt{2K_i}\right)}{2}}\Theta \geq \frac{\log(\epsilon_i)}{e^{\phi(\sqrt{2K_i})}},
\end{align}
where $\Theta\triangleq\int_{h}^{d_{\max}} {(2/L^2)d^{{\varphi\left(\sqrt{2K_i}\right)\alpha}/{2}+1}}  \text{d}d= {2\left(d_{\max}^{\frac{\varphi\left(\sqrt{2K_i}\right)\alpha}{2}+2} -h^{\frac{\varphi\left(\sqrt{2K_i}\right)\alpha}{2}+2}\right)}/{(L^2\;\left({{\varphi\left(\sqrt{2K_i}\right)\alpha}/{2}+2}\right))}.$
 
In order to fulfill the lower bound of rate coverage probability, minimum resources will be required, which results into the following relation:
\begin{align}
  \left( 2(K_i+1)\left(2^{ \frac{\eta_{th_i}}{\tau_i}}-1\right)\frac{\sigma^2}{\mu_i P_i}\right)={\left(\frac{- \log(\epsilon_i)}{e^{\phi(\sqrt{2K_i})}\Theta}\right)}^{\frac{2}{\varphi\left(\sqrt{2K_i}\right)}}.
\end{align}

From above equation we get:
\begin{align}\label{eq:Pi_to_ti}
P_i= V_i\left(2^{ \frac{\eta_{th_i}}{\tau_i}}-1\right).
\end{align}
where $V_i=  \frac{2(K_i+1)\sigma^2}{\mu_i {\left(\frac{ -\log(\epsilon_i)}{e^{\phi(\sqrt{2K_i})}\Theta}\right)}^{\frac{2}{\varphi\left(\sqrt{2K_i}\right)}}}$. Here \eqref{eq:Pi_to_ti} represents  closed-form expression for one optimization variable in terms of other variable. By doing so,  problem $\mathcal{P}_2$  is converted into one  dimension problem and can be solved by solving only one equation with very low  complexity (Section \ref{JOP}). 

\Remark
If we relax the assumption of common power source to all the users by considering that each user $U_i, \forall i \in I$  has independent power source with maximum power $P_{i,\max}$, and from \eqref{eq:Pi_to_ti} we get $\tau_i ={\eta_{th_i}}/{\log_2\left(({P_i}/{ V_i})+1\right)}$, then it can be noticed that since $\tau_i$ is decreasing function of $P_i$, each user  should use its maximum available power $P_{i,\max}$ to full-fill its service requirement  with minimum  time  $\tau_i={\eta_{th_i}}/{\log_2\left({(P_{i,\max}}/{ V_i})+1\right)},  \forall i \in I$, such that remaining time can be utilized to serve extra users  to maximize  UAV utility.

Now, to find the optimal $N^*$, we use the iterative search method on $\sum_{i=1}^{N} \tau_i \leq 1$ for $N$ in range $\{N_{lb}, N_{ub}\}$ such that $N^*$ is maximum possible $N$ where $\sum_{i=1}^{N^*} \tau_i \leq 1$ and  $\sum_{i=1}^{N^*+1} \tau_i > 1$.

\subsubsection{Variable Transformation in High SNR Case}
In this case, using \eqref{eq:highSNR}, 
 the constraint $(C1)$  can be written as $1-\mathcal{M}_i \Theta \geq \epsilon_i$, where $\mathcal{M}_i=e^{\phi(\sqrt{2K_i})}\; \left( 2(K_i+1)\left(2^{ \frac{\eta_{th_i}}{\tau_i}}-1\right)\frac{\sigma^2}{\mu_i P_i}\right)^{\frac{\varphi\left(\sqrt{2K_i}\right)}{2}}$. Using this constraint and $\widehat{V}_i=  \frac{2(K_i+1)\sigma^2}{\mu_i {\left(\frac{ (1-\epsilon_i)}{e^{\phi(\sqrt{2K_i})}\Theta}\right)}^{\frac{2}{\varphi\left(\sqrt{2K_i}\right)}}}$, we can express power in terms of time as:
 
\begin{align}\label{eq:Pi_to_ti_highSNR}
 P_i= \left( \frac{2(K_i+1)\left(2^{ \frac{\eta_{th_i}}{\tau_i}}-1\right)\sigma^2}{\mu_i {\left(\frac{(1-\epsilon_i)}{e^{\phi(\sqrt{2K_i})}\Theta}\right)}^{\frac{2}{\varphi\left(\sqrt{2K_i}\right)}}}\right)
= \widehat{V}_i\left(2^{ \frac{\eta_{th_i}}{\tau_i}}-1\right).
\end{align} 

\subsubsection{Variable Transformation in Dominant LoS Scenario}
In this case, using \eqref{eq:COV_HR}, the constraint $(C1)$ on rate-coverage probability can be written as $\frac{(P_ig_i)^{\frac{2}{\alpha}}L^{-2}}{(2^{\frac{\eta_{th_i}}{\tau_i}}-1)^{\frac{2}{\alpha}}(\sigma^2)^{\frac{2}{\alpha}}}-\frac{h^2}{L^2}\geq \epsilon_i$. From this constraint along with ${V}^K_i=  \frac{(\epsilon_i L^2+h^2)^{\frac{\alpha}{2}}\sigma^2}{g_i}$, we can express power in terms of time variable as follows:
\begin{align}\label{eq:Pi_to_ti_highSNR}
 P_i= \left( (\epsilon_i L^2+h^2)^{\frac{\alpha}{2}}\left(2^{ \frac{\eta_{th_i}}{\tau_i}}-1\right)\frac{\sigma^2}{g_i}\right)
= {V}^K_i\left(2^{ \frac{\eta_{th_i}}{\tau_i}}-1\right).
\end{align} 
\subsubsection{Low Complexity Design for Joint Optimal Solution}\label{opt_lambert} \label{low_complexity}
Here, we focus on the energy $e_i=P_i\tau_i$, allocated to single user $U_i$  and check the convexity of this term.
Using \eqref{eq:Pi_to_ti}, let us consider $e_i=P_i\tau_i=V_i\left(2^{ {\eta_{th_i}}/{\tau_i}}-1\right)\tau_i$; note that here in place of $V_i$, we can use $\widehat{V}_i$ and ${V}^K_i$  for the high SNR case and dominate LoS scenario, respectively. Taking its first and second derivative we get, $\frac{de_i}{d\tau_i}=V_i\left(2^{ \frac{\eta_{th_i}}{\tau_i}}-1-2^{ \frac{\eta_{th_i}}{\tau_i}} \frac{\eta_{th_i}\ln(2)}{\tau_i}\right)$ and  $\frac{d^2e_i}{d\tau_i^2}=V_i\left(2^{ \frac{\eta_{th_i}}{\tau_i}} \frac{( \eta_{th_i}\ln(2))^2}{(\tau_i)^3}\right)$, respectively. Since the second derivative is positive, the term $e_i$ is convex function of $\tau_i$ and has single root to minimize the term $e_i$ that can be obtained by equating the first derivative $\frac{de_i}{d\tau_i}$ to zero, i.e., $\left(2^{ \frac{\eta_{th_i}}{\tau_i}}-1-2^{ \frac{\eta_{th_i}}{\tau_i}} \frac{ \eta_{th_i}\ln(2)}{\tau_i}\right)=0$, which after rearrangement reduces to: $2^{{\eta_{th_i}}/{\tau_i}}\left(-\ln(2){\eta_{th_i}}/{\tau_i}+1\right)=1$. On solving this for $\tau_i$, by using a property that solution for $2^x(ax+b)=c$ is given by $x={\bold{W\left(\ln(2)\frac{c}{a}2^{\frac{b}{a}}\right)}}/{\ln(2)}-{b}/{a}$, we  obtain the solution for $\frac{\eta_{th_i}}{\tau_i}={\bold{W\left(-2^{\frac{1}{\ln(2)}}\right)}}/{\ln(2)}+\frac{1}{\ln(2)}$, in the form of Lambert function $\bold{W}(\cdot)$, with $a=-\ln(2)$, $b=1$ and $c=1$. Since, $\bold{W\left(-2^{{-1}/{\ln(2)}}\right)}=-1$, we get ${\eta_{th_i}}/{\tau_i}=0$, i.e. $\tau_i=\infty$. we are getting the root on right most corner for unbounded $\tau_i$, i.e. $e_i$ is the convex decreasing function of $\tau_i$. The physical interpretation of this result is that the energy consumption  $e_i$ can be reduced  by allocating the maximum possible time to user $U_i$. Since, power allocation $P_i$ is decreasing in $\tau_i$ with relation $P_i=V_i\left(2^{ {\eta_{th_i}}/{\tau_i}}-1\right)$, the maximum possible time allocation results into minimum power allocation to user $U_i$ to minimize the energy consumption in fulfill its service demand. 

\section{Rate Coverage Probability Analysis for Air-to-Ground Channel Modeling} 
Here, first we discuss the air to ground channel modeling and then analyze the rate coverage
probability. At last, the resource allocation method to the users for this case is discussed.
\subsection{Air-to-Ground  Channel Modeling}
One common approach for  air-to-ground channel modeling is to consider the LOS and NLoS separately along with their different occurrence probabilities \cite{Hourani}. Note that for NLoS link, path loss is higher than that in LoS link due to the shadowing effect and reflection from obstacles. 
Hence, some excessive path loss value is assigned to NLoS link along with the free space propagation loss. Furthermore, UAV-to-ground-user and ground-user-to-UAV link characteristics are assumed to be same. With this consideration, the received signal-to-noise ratio (SNR) $\Upsilon_i$ at UAV for the user $U_i$ can be written as  \cite{Moza_D2D}:
\setcounter{equation}{30}
\begin{equation}\label{eq:SNR_LOS}
\Upsilon_i \triangleq \left\{\begin{array}{ll}
{P_i d_i^{-\alpha}}/{\sigma^2}, \; \forall i\in I,  \quad \text{LoS link},  \\
 {\varsigma P_i d_i^{-\alpha}}/{\sigma^2}, \; \forall i\in I,  \quad \text{NLoS link}, 
 \end{array}\right.
\end{equation} 
where $P_i$ is the transmission power of user $U_i$, $\sigma^2$ denotes the AWGN, $d_i=\sqrt{r_i^2+h^2}$ is the distance of UAV from User $U_i$, and $\varsigma$ denotes the mean additional attenuation factor for the NLoS link. The probability of LoS link between  $U_i$ and UAV depends upon the elevation angle $\theta_i=\frac{180}{\pi}\times \sin^{-1}\frac{h}{d_i}$, density and height of buildings, and environment. The LoS probability $p_{LoS}$ is written as \cite{Hourani}:
\begin{equation}
p_{LoS}={1}/{(1+C\exp(-B[\theta-C]))},
\end{equation} 
where $C$ and $B$ are constant that depend on the environment (rural, urban, dense urban).
 The probability of NLoS link is $p_{NLoS}=1-p_{LoS}$.
In next subsection, the analysis of rate coverage probability is presented.

\subsection{Rate Coverage Probability Analysis} Using the Shannon's capacity formula and SNR representation in \eqref{eq:SNR_LOS}, the spectral efficiency for user $U_i$ is given by:
\begin{equation}\label{eq:rate_los}
\tilde{\eta_i}= \tau_i\log_2(1+\Upsilon_i), \forall i\in I,
\end{equation}
Where $\tau_i$ is normalized fraction of time allocated from block duration $T$ to user $U_i$, $\forall i\in I$.
Now, a randomly chosen tagged user $U_i$, can be considered under rate coverage if the assigned data rate to that user is greater than its defined threshold $\eta_{th_i}$. Thus, rate coverage probability can be expressed mathematically as follows:
\begin{align}\label{eq:rate_cov_LOS}
\tilde{\mathcal{P}}_{cov_i} &= \mathbb{P}[\tau_i \log_2(1+\Upsilon_i)\geq \eta_{th_i}] = \mathbb{P}\left[\Upsilon_i\geq \left(2^{\frac{\eta_{th_i}}{\tau_i}}-1\right)\right] \nonumber\\
&\overset{(a_1)} {=}p_{LoS}\mathbb{P}[\Upsilon_i\geq w_1|\text{LoS}]+p_{NLoS}\mathbb{P}[\Upsilon_i\geq w_1|\text{NLoS}]\nonumber\\
&\overset{(a_2)} {=}p_{LoS}\mathbb{P}\left[r_i\leq \sqrt{({P_i}/({\sigma^2 w_1}))^{2/\alpha}-h^2}\right]\nonumber\\
&\hspace{0.2in}+p_{NLoS}\mathbb{P}\left[r_i\leq \sqrt{({\varsigma P_i}/({\sigma^2 w_1}))^{2/\alpha}-h^2}\right]\nonumber\\
&\overset{(a_3)} {=} \int_0^{ \sqrt{({P_i}/({\sigma^2 w_1}))^{2/\alpha}-h^2} }\frac{p_{LoS}.2r}{L^2} dr\nonumber\\
&\hspace{0.2in}+ \int_0^{ \sqrt{({\varsigma P_i}/({\sigma^2 w_1}))^{2/\alpha}-h^2} } \frac{p_{NLoS}. 2r}{L^2} dr\nonumber\\
&\overset{(a_4)} {=}\frac{{({\varsigma P_i}/({\sigma^2 w_1}))^{2/\alpha}-h^2}}{L^2}\nonumber\\
&\hspace{0.2in}+\int_{\sqrt{({\varsigma P_i}/({\sigma^2 w_1}))^{2/\alpha}-h^2}} ^{ \sqrt{({P_i}/({\sigma^2 w_1}))^{2/\alpha}-h^2} }\frac{p_{LoS}.2r}{L^2} dr,
\end{align}
 where $w_1=2^{\frac{\eta_{th_i}}{\tau_i}}-1 $. Equality $(a_1)$ is obtained by considering the probability of occurrence of LoS and NLoS link. Whereas $(a_2)$ uses \eqref{eq:SNR_LOS}, $(a_3)$ comes by considering the  PDF, $f_R(r)=\frac{2r}{L^2},  0\leq r \leq L$, of horizontal distances of uniformly distributed user from the center of circular field of radius $L$, and $(a_4)$ comes by substituting $p_{NLoS}=1-p_{LoS}$.
 In the next subsection, resource allocation method is discussed to solve the utility maximization problem $\mathcal{P}_2$.
\subsection{ Power $P_i$ and Time $\tau_i$ Resource Allocation} Since the rate coverage probability derived in \eqref{eq:rate_cov_LOS} can not be expressed in closed-form, the variable transformation method can not be applied in this case to obtain the optimal time and power resource allocation to the users. Therefore, while considering the air-to ground channel modeling, the optimization problem $\mathcal{P}_2$ has to be solved numerically, while taking care of total time and power budget constraints. In order to reduce the solution complexity, the resource allocation to the users is done in distributed manner, such that the users' rate coverage demand can be fulfilled with minimum possible energy resources and maximum possible users can be served, which in turn is equivalent to the UAV utility maximization.

\section{Applicability of the proposed methodology in mobile UAV scenario}
Let, the UAV is following a trajectory with same initial and final points and $\mathcal{T}$ is total time to complete the trajectory by the UAV with maximum speed $V_{max}$. For ease of exposition, the total period $\mathcal{T}$ is divided in $n_t$ slots. The elemental slot
length $ \delta t= \frac{\mathcal{T}}{n_t}$ is chosen to be sufficiently small such that the UAV’s location is considered as approximately unchanged \cite{QURUI2}. To guarantee a certain accuracy with this approximation, the ratio of $S_{\max}=V_{\max}\delta t$ and altitude $h$ of the UAV can
be restricted below a threshold, i.e. $ \frac{V_{\max}\delta t}{h}\leq \varepsilon_{\max}$, where $\varepsilon_{\max}$ is the given threshold. Thus, the minimum number of time slots required for achieving  desired accuracy is given by:
\begin{equation}
n_t \geq {V_{\max} \mathcal{T}}/{(h\; \varepsilon_{\max})}.
\end{equation}
Now, we  equate  consider block time $T$ to the elemental slot time $\delta t$ and use the Algorithm $1$ for jointly-optimal power and time resource allocation to the users under service. Since, the location of the UAV changes, there is need to periodically recalculate the distances of users from the UAV and also reallocation of resource is done in every elemental slot $T=\delta t$.

Note that,  here to discuss the applicability of the proposed methodology in mobile UAV scenario, we have considered optimal resource allocation over different time slots of total flight period. However, it may not be the best solution in mobile UAV scenario. The optimization across the total flight period along with trajectory optimization in mobile UAV scenario is out of the scope of this work.

\bibliographystyle{IEEEtran}

\end{document}